\documentclass[showpacs,aps, twocolumn]{revtex4-2}
\usepackage{epsfig}
\usepackage{graphicx}
\usepackage{amsmath,amssymb,amsfonts}
\usepackage{array}
\usepackage{url}
\usepackage[usenames,dvipsnames]{color}
\usepackage[colorlinks=true,linkcolor=magenta, citecolor = blue]{hyperref}
\usepackage{multirow}
\usepackage{float}
\usepackage{lineno}
\usepackage{xspace}
\usepackage{ulem}
\usepackage{lipsum}

\begin{document}

\title{Exploring  QGP-like phenomena with charmonia in $p+p$ collisions at $\sqrt{s} = 13$ TeV }

\author{Captain R. Singh$^1$}
\email{captainriturajsingh@gmail.com}

\author{Partha Bagchi$^{2,3}$}
\email{parphy85@gmail.com}

\author{Raghunath Sahoo$^1$}
\email{Raghunath.Sahoo@cern.ch}

\author{Jan-e Alam$^4$}
\email{janephysics1996@gmail.com}

\affiliation{$^1$Department of Physics, Indian Institute of Technology Indore, Simrol, Indore-453552, India}
\affiliation{$^2$School of Physical Sciences, National Institute of Science Education and Research, Jatni, Odisha
-752050, India}
\affiliation{$^3$Physics Department, Marwari College, Kishanganj (A constituent unit of Purnea University),
Bihar–855107, India}
\affiliation{$^4$Murshidabad University, Berhampore, Murshidabad-742101, India}

\begin{abstract}

In ultrarelativistic collisions of nuclei at the Large Hadron Collider, the created QCD environment rapidly
changes, leading to a
nonadiabatic evolution of the quantum states involved. Considering this, we first examine the preequilibrium
state of QCD matter
and its effect on the initially produced charmonium using a temperature-independent Hamiltonian. As the QCD
matter reaches
local thermal equilibrium, this Hamiltonian transforms to its finite temperature counterpart. To model the
preequilibrium stage, we use the bottom-up thermalization approach to determine the effective temperature of
the QCD
matter, followed by a Gubser-type expansion for the thermalized medium. Additionally, we consider collisional
damping,
gluonic dissociation, and regeneration mechanisms, which specifically modify the charmonium yield in the
thermalized
medium. Mainly, the gluonic dissociation and collisional damping cause a reduction in the yield conversely,
regeneration through gluonic deexcitation enhances the yield of charmonium. Further, we explore the combined
effects of these
mechanisms on the collective yield of charmonium states with transverse momentum ($p_{\rm T}$) and event
multiplicity
in the proton-proton collisions at $\sqrt{s} = 13$ TeV. Based on our findings, we contend that the combined
effects of
these mechanisms can serve as a robust probe for determining the possible existence of a thermalized QCD
medium in such
a small collision system.

\pacs{}
\end{abstract}
\date{\today}
\maketitle

\section{Introduction}
\label{intro}
The suppression of quarkonia has been proposed as an efficient probe for the creation of the transient phase of
quark-gluon plasma (QGP) in heavy-ion collisions~\cite{Matsui:1986dk, NA50:2000brc}. In QGP-like scenarios,
quarkonia
suppression arises from the breaking of the heavy quark-antiquark pair ($Q-\bar{Q}$) and the screening of the
QCD  potential, the transition from color neutral to a colored state~\cite{Matsui:1986dk, nendzig,
Singh:2015eta}. In the heavy-ion collisions, even in the absence of a QGP-like medium, the quarkonia
production itself
gets suppressed to a certain extent  due to the presence of a nuclear environment in the colliding ions, such
phenomena
are
incorporated through the cold nuclear matter (CNM) effects~\cite{vogt}. However, the CNM effect and the QGP
effect were separately
unable to explain the experimental data of  quarkonia suppression from heavy-ion collisions at RHIC and LHC
energies. The inclusion of these two effects jointly helped to explain the quarkonia suppression data
qualitatively in Au$-$Au, Pb$-$Pb, $p-$Pb collisions at RHIC and LHC energies~\cite{Singh:2015eta,
Singh:2018wdt, Ganesh:2016kug, Singh:2021evv, Hatwar:2020esf}. In these studies, gluonic dissociation,
collisional
damping, and color screening are the main effects reducing the effective yield of the quarkonia in the QGP
medium.
Under CNM effects, authors ~\cite{vogt} considered the shadowing effect, which modifies the initial production
of the quarkonia in
heavy-ion collisions. Besides the suppression or mechanism of yield reduction, an enhancement in the yield due
to
secondary production or regeneration of quarkonia in the QGP medium is also considered. Meanwhile, quarkonia
suppression in heavy-ion collisions is an interplay of various phenomena including cold and hot nuclear matter
effects.\\

The quarkonium suppression in heavy-ion collisions is studied by considering results from $p+p$ collisions as a
baseline~\cite{NA50:2000brc,rhic,exp1,exp2}. The $p+p$ collision is used as a benchmark because it is assumed
that
such collisions lack the nuclear environment and are also unable to achieve the critical conditions to create a
thermal QCD medium. But over the decades, a significant increase in the center of mass collision energies at
the LHC
has been achieved and results from $p+p$ collisions have changed this perception.
The data from high-multiplicity $p+p$ collisions at $\sqrt{s}$ = 7 TeV and 13 TeV have shown the phenomena
which resemble
the conditions of heavy ion collision~\cite{nature, atlas2, atlas3,cms2}. However, the existence of QGP in
such a small system
is still unclear and requires more investigation. In this direction, experimental data of the normalized
charmonium
yield observed in $p+p$ collisions have been quantitatively explained using the unified model of quarkonia
suppression
(UMQS)~\cite{Singh:2021evv}. The UMQS model is based on the QGP phenomenology and it was successful in
explaining the
quarkonia suppression in A$-$A and $p-$A collisions at various center-of-mass energies. And in the given
conditions it predicts a QGP-like scenario in the $p+p$ collisions at the LHC energy.\\

The quarkonia suppression in heavy-ion collisions due to the production of  QGP was first proposed by Matsui
and Satz
\cite{Matsui:1986dk} based on the color screening mechanism. In their work, it was considered that when the
temperature of the medium exceeds the dissociation/melting temperature of the quarkonium, the quark-antiquark
potential gets screened, resulting in the suppression of these states. It infers that quarkonia have adequate
time to
adjust to the evolving medium, thereby undergoing adiabatic evolution. The adiabatic evolution involves gradual
changes, enabling the system to adapt its configuration over time. As a consequence, the change in the
Hamiltonian of
the system must occur slowly to prevent transitions to different eigenstates. However, such conditions may not
be satisfied in small
collision systems like $p+p$, where the temperature of the fireball is extremely high but the system size is
very small.
Such a system is expected to cool down rapidly and consequently, the rapid evolution of the plasma dynamics can
challenge the conditions required for adiabatic evolution. This necessitates a theoretical framework that
accounts for
nonadiabatic evolution, where quarkonia states make a transition to other bound states or continuum states due
to rapid
changes in temperature \cite{Dutta:2012nw, Bagchi:2014jya, Dutta:2019ntj, Bagchi:2018auv, Boyd:2019arx,
Atreya:2014sea}.
A similar scenario may exist in noncentral heavy-ion collisions, where a transient magnetic field can
contribute to
the nonadiabatic evolution of medium viz quarkonia~\cite{Bagchi:2023jjk, Iwasaki:2021nrz, Iwasaki:2021kms,
Guo:2015nsa}. As the evolution of quarkonia depends on the QGP lifetime, particularly the temperature decay
rate and
the initial temperature of the medium, a rapid decrease in the temperature may not allow sufficient time for
quarkonia to dissociate, even if the initial temperature surpasses the dissociation threshold. By extending
the concept
of adiabatic evolution, it is argued that the effective temperature determines the fate of quark-antiquark
bound states
in $p+p$ collisions. If the effective temperature exceeds the dissociation temperature, bound states dissolve;
otherwise, dissociation is minimal. As discussed, in the $p+p$ collisions, the rapid temperature reduction can
abbreviate the lifespan of the deconfined QCD medium, leading to abrupt alterations in the Hamiltonian of the
quarkonia. Consequently, it permits nonadiabatic evolution to take place.\\

In this study, we consider the Gubser-like expansion of the medium created in ultrarelativistic $p+p$
collision, which
predicts that the thermalized medium gets exhausted in a very brief time, say $\le 1$ fm. It allows us to
delve into the
suppression of charmonium by incorporating nonadiabatic evolution, showcasing how it can extend the
persistence of
quark-antiquark bound states even amidst heightened multiplicities \cite{Bagchi:2023vfv}. Following this, we
study
the yield modification of the $J/\psi$, $\chi_c$ and $\psi$(2S) in $p+p$ collision via incorporating
nonadiabatic
evolution of quarkonia along with the collision damping, gluonic dissociation, and regeneration
mechanisms as the QGP effects~\cite{Singh:2015eta}.\\

The paper is organized as follows. In Sec.~\ref{formalism}, we discuss the dynamics of the fireball by
modeling temperature evolution in the prehydrodynamic or preequilibrium phase followed by Gubser flow for
the thermalized/hydrodynamic phase. In this section, we also discuss the modification in the temperature in the
particle rest frame caused by the relativistic doppler shift (RDS). Next, Sec.~\ref{yield}, incorporates the
dissociation probability of quark-antiquark bound states as well as transitions to other states under
nonadiabatic
evolution using time-dependent perturbation theory. Further, it briefly describes the regeneration of
charmonium during
QGP evolution. It takes us to the next Sec.~\ref{results}, which  presents the main outcomes of the study,
demonstrating the yield modification of different charmonium states against charged particle multiplicity
($\frac{dN_{ch}}{d\eta}$) and transverse momentum ($p_{T}$). We also observe the modification of the $\chi_c$
and $\psi$(2S) yields with respect to $J/\psi$ in terms of double ratio as well as number ratios.  Finally,
Sec.~\ref{summary} concludes and summarizes the results, providing an outlook on future research.

\section{System Dynamics}
\label{formalism}
This section is divided into three main parts. Firstly, we will discuss the solution of the time-dependent
Schr$\ddot{o}$dinger equation, focusing on the effects of rapid changes in potential on bound states.
Subsequently, we
explore the evolution of temperature during the preequilibrium stage. Lastly, we analyze the temperature
evolution in
the late stage after hadronization or during the near-equilibrium stage.\\

\subsection{Preequilibrium Kinematics}

The key quantity that controls the evolution of wave function and specifically modifies the survival
probability is the
Hamiltonian, which carries the temporal dependence originating from the time dependence of the temperature.
Modeling the time evolution of temperature is nontrivial for the entire evolution of the plasma in heavy ion
collisions. Fortunately, hydrodynamic evolution plays a crucial role in the space-time evolution of the QCD
medium
after the partonic medium thermalizes. The hydrodynamics successfully describes the bulk evolution of the
medium. Therefore,
we can choose a  hydrodynamic model to study the temperature evolution which governs the evolution of the
Hamiltonian.
However, to model the  preequilibrium stages, one may rely on the effective QCD kinetic theory description as
discussed within the
framework of bottom-up thermalization~\cite{Kurkela:2018oqw,Kurkela:2018xxd,Kurkela:2018vqr}. Qualitatively,
in this approach,
it has been argued that in nonexpanding systems, gauge bosons (gluons here) can rapidly achieve equilibrium
(kinetic) among
themselves, followed by the equilibration of the fermions. On the other hand, if the system undergoes rapid
longitudinal expansion,
partons may remain out of equilibrium, but the system can be effectively described by fluid dynamics. Without
going
into the details of the model, we consider the following ansatz for the proper time evolution of the
\textit{pseudo}
temperature ($T_{\rm{p}}$). Note that it is a pseudotemperature because the temperature is strictly defined
only in
equilibrium:

\begin{equation}
\frac{T_{\rm{p}}}{T_{\rm{Hydro}}}=\left(\frac{\tau}{\tau_{\rm{Hydro}}}\right)^{\frac{1}{7}\frac{\alpha-1}{(\
alpha+3)}}
\label{equ9}
\end{equation}

The physical picture that prompts us to explore the above scaling is that the initial out-of-equilibrium
partons
scatter with each other to achieve kinetic/thermal equilibrium. We also identify the thermalization timescale
as the
time when we can apply the hydrodynamic description ($\tau_{\rm{Hydro}}$). In principle, all these different
timescales can form a hierarchy, but we expect that if thermalization is achieved very fast, then the
difference between
different scales may not be too large, not affecting the system dynamics significantly. The parameter $\alpha$
enters
the above equation because the pseudo temperature can be defined through the $\alpha$th moment of the
fermionic or
bosonic distribution function~\cite{Kurkela:2018oqw}. Physically, the parameter $\alpha$ determines how fast
the system
achieves hydronization (onset of hydrodynamic description) or thermalization. In the subsequent discussion, we
appropriately choose $T_{\rm{Hydro}}$, $\tau_{\rm{Hydro}}$, and $\alpha$ to model the preequilibrium
dynamics. Instead
of going into the microscopic description, naively, one can also assume the temperature starts at zero at some
initial
time and increases linearly until it reaches a value $T_{\rm{Hydro}}$ at time $\tau_{\rm{Hydro}}$.\\

\subsection{Thermal evolution with Gubser flow}
Once we have a description of the preequilibrium pseudo temperature that also quantifies the preequilibrium
dynamics of the Hamiltonian, we can look into the temperature evolution due to the flow dynamics. To solve
the hydrodynamic equations, defining both the initial conditions and the equation of state  is essential.
In the
absence of a first-principle method for estimating the initial temperature ($T_0 = T_{\rm{Hydro}}$), the
following
relation has been employed in this study to constrain $T_0$ using available data~\cite{hwa3}:

\begin{equation}
T_{0} = \left[\frac{90}{g_{k} 4\pi^{2}}C^{\prime}\frac{1}{A_T\tau_0}1.5\frac{dN_{ch}}{dy}\right]^{1/3}
 \label{t0}
\end{equation}
where $A_T$ is the transverse area of the system obtained using the IP-Glasma model~\cite{McLerran:2013oju},
and $g_k$ is
the statistical degeneracy of the QGP phase. In Eq.~(\ref{t0}), $C^{\prime} = \frac{2\pi^{4}}{45\zeta(3)}
\approx 3.6$.
Additionally, we assume $\frac{dN_{ch}}{dy} \cong \frac{dN_{ch}}{d\eta}$, which holds true in the massless
limit. Given
the lack of a first-principle approach to determine thermalization time ($\tau_{0} = \tau_{\rm{Hydro}}$), it is
reasonable to hypothesize that the thermalization
time decreases with increasing center-of-mass collision energy, i.e., $\tau_{0}\propto
1/\sqrt{s}$~\cite{hwa2,hwa3}. In
this study, we assume $\tau_{0} = 0.1$ fm for $pp$ collisions at $\sqrt{s} = 13$ TeV.\\

Notably, in $p+p$ collisions, the size of the produced medium is expected to be relatively small, and the
maximum
size for high-multiplicity $p+p$ collisions can approximately be $1.5$ fm~\cite{McLerran:2013oju}.
Consequently, transverse
expansion must be addressed. To account for the transverse
expansion of the system in this calculation, we examine the Gubser flow, first explored by Gubser and Yarom
\cite{Gubser:2010ze, Gubser:2010ui}. This approach combines a ``boost-invariant" longitudinal flow, akin to
the Bjorken
flow, with consideration for transverse flow. The evolution of thermodynamic quantities, including energy
density
($\epsilon$) and shear stress ($\pi$), within the framework of Gubser flow with third-order viscous
corrections, is
detailed in \cite{Chattopadhyay:2018apf, Dash:2020zqx}.\\

\begin{eqnarray}
 \frac{d\hat\epsilon}{d\rho} &=& -\left(\frac{8}{3}\hat\epsilon-\hat{\pi} \right)\tanh(\rho)
\label{eq:e_evolution} \\
 \frac{d\hat\pi}{d\rho} &=&
-\frac{\hat\pi}{\hat\tau_{\pi}}+\tanh(\rho)\left(\frac{4}{3}\hat\beta_{\pi}-\hat\lambda\hat\pi-
\hat\chi\frac{\hat{\pi}^2}{\hat\beta_{\pi}}\right)\label{eq:pi_evolution}
\end{eqnarray}

The dimensionless quantities, $\hat\epsilon$ and $\hat\pi$, are expressed as $\hat\epsilon= \hat{T}^4 =
\epsilon \tau^4
= 3\hat{P}$ and $\hat\pi=\pi\tau^4$ where $\tau$ is the proper time and $\hat{T}$ is related to temperature.
The
parameters are chosen \cite{Chattopadhyay:2018apf} as $\epsilon = \frac{3}{\pi^2} T^4$, $\hat\tau_\pi
(=c/\hat{T})$ is
related to relaxation time, where $c=5\frac{\eta}{s}$, $\hat\beta_\pi= 4\hat{P}/5$, $\hat{\lambda}=46/21$ and
the third-order correction parameter $\hat\chi=72/245$.\\

The conformal time $\rho$ can be written as
\begin{equation}
    \rho = -\sinh^{-1}\left(\frac{1-q^2\tau^2+q^2x_T^2}{2q\tau}\right)
\end{equation}
where $q$ is an arbitrary energy scale, which is related to the transverse size of the medium ($r_T$) like
$q=\frac{1}{r_T}$, $x_T$ is the position in the transverse plane. One can retrieve the Bjorken flow solution
by taking
the limit $r_T\rightarrow \infty$ or $q\rightarrow 0$. One can also use the $(3+1)$-dimensional hydrodynamic
description for a more accurate description of nonboost invariant flow with nontrivial rapidity dependence.
But considering the
possible boost invariance in ultrarelativistic collisions, we restrict ourselves to the analytically solvable
hydrodynamic description with transverse expansion.\\

Previously, the effects of the transverse expansion on the temperature
evolution given by Eqs.(\ref{eq:e_evolution}) and (\ref{eq:pi_evolution}) with initial
conditions $ T = T_{\rm{Hydro}} = 350$ MeV and $\hat\pi = \frac{4}{3}\hat\beta_{\pi}\hat\tau_{\pi}$ at $\tau =
\tau_{\rm{Hydro}} = 0.3$ fm for various system sizes ($r_T$) was demonstrated in Ref.~\cite{Bagchi:2023vfv}.
The results indicate that, as $r_T$ increases, the
lifetime of the QGP increases. At sufficiently large $r_T$, the variation of temperature
($T$)
with proper time ($\tau$) for Gubser flow closely resembles that of Bjorken flow. The variation of temperature
with system
size clearly indicates that for small systems, the time evolution of the system can be rapid as compared to
large
systems, allowing us to explore the scenario of nonadiabatic evolution.\\


 \subsection{In-medium implicit temperature for quarkonia}

Heavy quarkonia like charmonia do not attain thermalization with the medium, and such mesons have a different
velocity than the medium. The velocities of the medium and charmonium are denoted by $v_{m}$ and
$v_{J/\psi(nl)}$,
respectively. This lack of integration between charmonium and the surrounding medium gives rise to acquiring an
effective temperature for charmonium. The effective temperature of charmonium in the medium is obtained using
the
RDS, which arises from the velocity difference between the charmonium and the
thermalized
QCD medium. The RDS leads to an angle-dependent effective temperature ($\rm T_{eff}$), expressed
as~\cite{PhysRevD.84.016008, PhysRevD.87.114005}:\\

\begin{equation}
\rm T_{\rm eff}(\theta,|v_{r}|) = \frac{T(\tau)\;\sqrt{1 - |v_{r}|^{2}}}{1 - |v_{r}|\;\cos \theta}
\label{tt}
\end{equation}
where $\theta$ is the angle between the
relative velocity $v_{r}$ and the direction of the free-flowing light partons. The $\rm T(\tau)$ in
Eq.~(\ref{tt}),
represents the medium cooling rate obtained using Gubser flow. For a very narrow region, 
$0 < \theta \le \pi/4$,  Eq.~(\ref{tt}) predicts the $\rm T_{\rm eff}$ larger than the medium temperature (T),
{\it i.e.}
$ \rm T_{\rm eff} > $T 
while elsewhere,
$ \rm T_{\rm eff} < $T. Now  for $ \rm T_{\rm eff} > $T, particles might dissociate while the medium
temperature may
not be large enough to induce such dissociation, such a situation seemed unphysical  as it implies
dissociation occurring
under subcritical conditions. To fix this issue, we averaged the $ \rm T_{\rm eff}$ over solid angle $\theta$,
which ensures that the integrated effective temperature is physically consistent with the actual thermal
environment experienced by the moving charmonium. This phenomena is supported  in
the Ref.~\cite{Singh2024}, where the angle average $ \rm T_{\rm eff}$  was shown to remain below T, preserving
causal consistency. Now, the angle-independent effective temperature is given as~\cite{Singh:2018wdt};

\begin{equation}
\rm T_{\rm eff}(\tau, p_{T}) = T(\tau)\;\frac{\sqrt{1 -
|v_{r}|^{2}}}{2\;|v_{r}|}\;\ln\Bigg[\;\frac{1 + |v_{r}|}{1 - |v_{r}|}\Bigg]
\label{teff}
\end{equation}
It should be noted that different observables might be sensitive to the different functional dependencies of
the temperature
average over angle, such as $<T>, <T^{2}>,$ etc. However, in the employed formulation, thermal decay widths are
propotional to T
and therefore in this context, the obtained angle averaged $ \rm T_{\rm eff}$ serves physical and practical
consistency.\\

\begin{figure}
\includegraphics[width=0.49\textwidth]{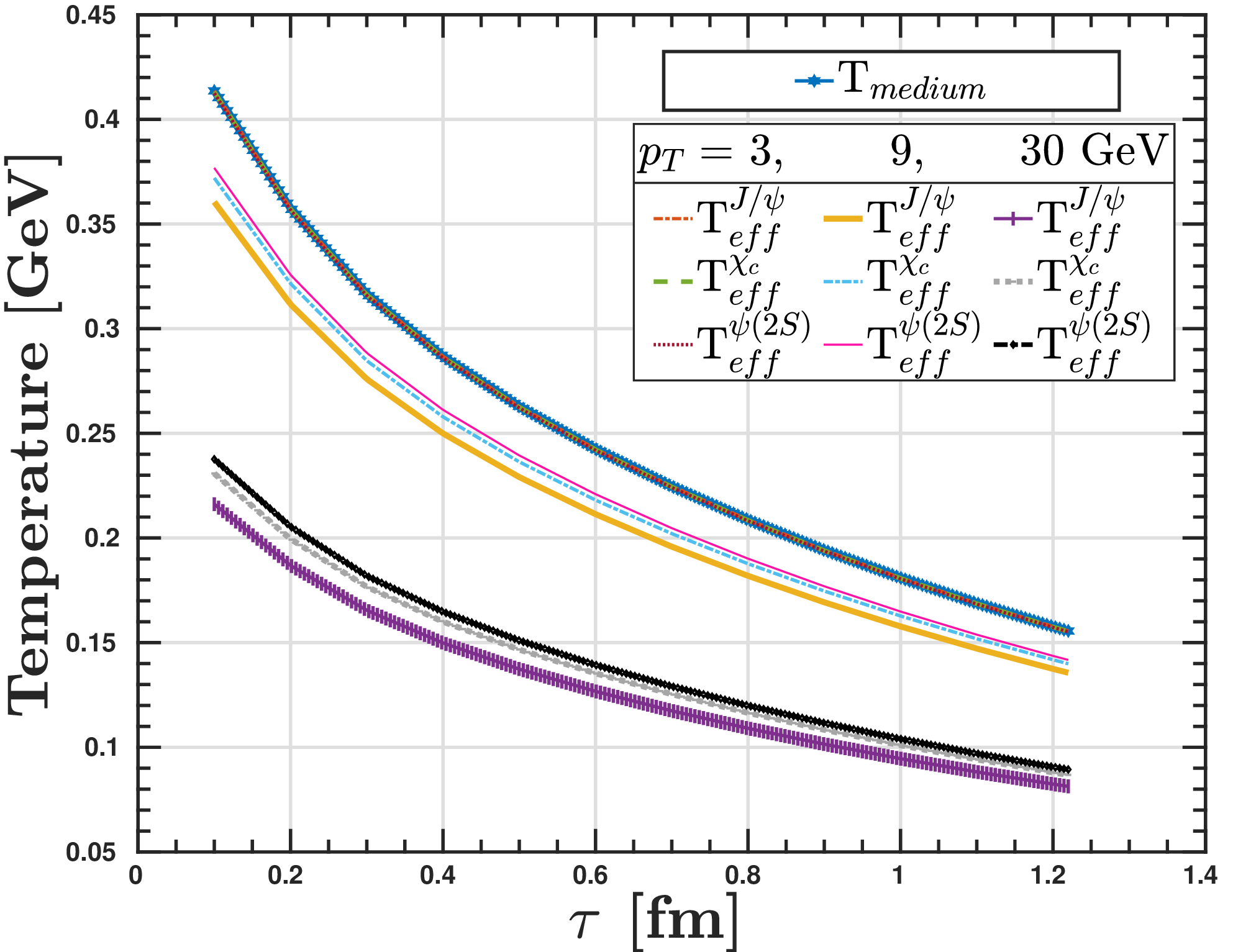}
\caption{(Color online) Medium temperature evolution with time ($\tau$) corresponding to Gubser flow is
represented
through $\rm T_{medium}$. Along with it, the effective temperature ($\rm T_{eff}$) respective to $J/\psi$,
$\chi_{c}$(1P)
and $\psi$(2S) for $p_{T} = $ 3, 9, 30 GeV are also shown here.}
\label{fig:Gflow}
\end{figure}

Now, using Eq.~(\ref{teff}), we have obtained the variation of the effective temperature experienced by
$J/\psi$, $\chi_c$
and $\psi$(2S) charmonium with $\tau$ and compare the results with medium temperature $\rm T_{medium}$ as shown
Fig.~\ref{fig:Gflow}. For $p_{T} \le 3$ GeV, all the charmonia resonances are found to be thermalized with
medium, as
$\rm T_{eff}$ corresponding to $J/\psi$, $\chi_c$ and $\psi$(2S) is almost same as $\rm T_{medium}$. While $\rm
T_{eff}$ obtained for $3<p_{T}\le9$ GeV comes out less than $\rm T_{medium}$, following the argument that
quarkonia
moving with high $p_{T}$ is incapable of being in thermal equilibrium with the medium. As $\psi$(2S) and
$\chi_c$
masses are higher than $J/\psi$, traverse through the medium with relatively slower speed, and consequently
feel
slightly higher temperatures at given $p_{T} > 3$ GeV. This mass ordering on the $\rm T_{eff}$ for $J/\psi$,
$\chi_c$ and $\psi$(2S) is preserved at $9<p_{T}\le30$ GeV while $\rm T_{eff}$ is further reduced respective
to the
$p_{T}$-range; $3<p_{T} \le9$ GeV. Our findings suggest that $\rm T_{eff}$ plays a crucial role in the
modification
of charmonium yield in the QGP medium.

\section{Yield Modification Mechanisms}
There are several phenomena related to cold nuclear matter and hot QCD matter that may influence the
production/suppression
of charmonium in heavy-ion collisions. However, unlike heavy-ion collisions, the nuclear environment is absent
in $p+p$ collisions.
Therefore, any changes in charmonium yield in these collisions can be attributed solely to the effects of a
hot partonic medium.
In the context of hot QCD matter effects, we have considered factors such as collisional damping, gluonic
dissociation, and the
regeneration of charmonium states through gluonic deexcitation. Additionally, we have formulated the
nonadiabatic approximation
for charmonium evolution and studied its impact on charmonium yield.

\label{yield}


\subsection{Collisional damping}
The collisional energy loss of charmonium within the QGP is characterized by ``collisional
damping." The charmonium dissociation due to this damping effect is evaluated using the complex singlet
potential. In
this study the singlet potential for the $c\bar{c}$ bound state in the QGP medium is defined as
follows~\cite{laine,
aber, Singh:2015eta}:\\\

\begin{multline}
V(r,m_D) = \frac{\sigma}{m_D}(1 - e^{-m_D\,r}) - \alpha_{eff} \left ( m_D
+ \frac{e^{-m_D\,r}}{r} \right )\\
- i\alpha_{\rm eff} {\rm T_{eff}} \int_0^\infty
\frac{2\,z\,dz}{(1+z^2)^2} \left ( 1 - \frac{\sin(m_D\,r\,z)}{m_D\,r\,z} \right)
\label{pot}
\end{multline}

In Eq.~(\ref{pot}), the first two terms on the right-hand side represent the string and Coulombic
contributions,
respectively. The third term is the imaginary component of the heavy-quark potential, which accounts for
collisional
damping. Here, $\sigma$ is the string tension for the $c\bar{c}$ bound state, and $m_{D}$ denotes the Debye
mass. The
running coupling constant at the hard scale, $\alpha_{s}^{T}$; $\alpha_{s}^{T} = \alpha_{s}(2\pi T)$. The
effective
coupling constant, $\alpha_{\rm eff}$, is defined at the soft scale as $\alpha_{s}^{s} = \alpha_{s}$, and is
expressed
as $\alpha_{\rm eff} = \frac{4}{3}\alpha_{s}^{s}$.\\

The collisional damping, $\Gamma_{c,nl}$, describes the decay of charmonium caused by the imaginary part of the
complex potential and it dominates at $m_D >> E$ (binding energy of charmonia)  and $m_{q} >>T >> 1/r$, here
$m_{q}$ is the bare mass of heavy quark, and  $r$ is
the size of  the bound state. The $\Gamma_{c,nl}$ is computed using first-order perturbation theory by
integrating the imaginary
part of the potential with the radial wave function:\\

\begin{equation}
\Gamma_{c,nl}(\tau,p_{T}) = \int[g_{nl}(r)^{\dagger} \left [ Im(V)\right] g_{nl}(r)]dr,
\label{cold}
\end{equation}
where $g_{nl}(r)$ is the charmonia singlet wave function. We have obtained the wave functions  by solving the
Schr\"{o}dinger equation for $J/\psi$, $\chi_{c}$(1P), $\psi$(2S).

\subsection{Gluonic dissociation}
 In QGP, quarkonium states can make a transition from a color singlet state to a color octet state via
absorption
of an E1 gluon (where E1 is the lowest electric mode for the spin-orbital wave function of gluons), and
gradually the color
octet state dissociates within the medium. The thermal decay width associated with this phenomenon is termed
gluonic
dissociation, it dominates at  $E >> m_D$ and $m_{q} >>1/r >>T$. The cross section for this process is given
by~\cite{Singh:2021evv}:

\begin{multline}
\sigma_{d,nl}(E_g) = \frac{\pi^2\alpha_s^u E_g}{N_c^2}\sqrt{\frac{{m_{c}}}{E_g
+ E_{nl}}}\\
\;\;\;\;\;\times \left(\frac{l|J_{nl}^{q,l-1}|^2 +
(l+1)|J_{nl}^{q,l+1}|^2}{2l+1} \right)
\end{multline}
where $m_c$ is the mass of the charm quark and $\alpha_s^u$ is the coupling constant, scaled as
$\alpha_{s}^{u} = \alpha_{s}(\alpha_{s}m_{c}^{2}/2)$. The $E_{nl}$ is the energy eigenvalue corresponding to
the
charmonium wave function, $g_{nl}(r)$. Here, $J_{nl}^{ql^{'}}$ is the probability density, derived using both
the
singlet and octet wave functions as follows;

\begin{equation}
 J_{nl}^{ql'} = \int_0^\infty dr\; r\; g^*_{nl}(r)\;h_{ql'}(r)
\end{equation}

The octet wave function $h_{ql'}(r)$ has been obtained by solving the Schr\"{o}dinger equation with the octet
potential
$V_{8} = \alpha_{\rm eff}/8r$. The value of $q$ is determined by using the conservation of energy, $q =
\sqrt{m_{c}(E_{g}+E_{nl})}$.\\

The gluonic dissociation rate, $\Gamma_{gd,nl}$, is calculated by taking the thermal average of the
dissociation
cross section~\cite{Singh:2021evv}:

\begin{equation}
\Gamma_{gd,nl}(\tau,p_{T},b) = \frac{g_d}{4\pi^2} \int_{0}^{\infty}
\int_{0}^{\pi} \frac{dp_g\,d\theta\,\sin\theta\,p_g^2
\sigma_{d,nl}(E_g)}{e^{ \{\frac{\gamma E_{g}}{\rm T_{eff}}(1 +
v_{J/\psi}\cos\theta)\}} - 1}
\label{glud}
\end{equation}
The $p_T$ is the transverse momentum of the charmonium, and $g_d$ represents the degeneracy factor of the
gluons.\\

Now taking the sum of the decay rates associated with collisional damping and gluonic dissociation, the
combined effect
is expressed in terms of the total decay width, given as;

\begin{equation}
 \Gamma_{D,nl} (\tau,p_{T}) = \Gamma_{c,nl} + \Gamma_{gd,nl}
\label{GmD}
\end{equation}

These two mechanisms dominate in different physical domains. However, collisional damping is the leading
dissociation factor, while
the gluonic dissociation has a marginal impact on quarkonium decay width. The gluonic dissociation increases
with the temperature
at a certain level and starts decreasing at high temperature. This decrease in decay width corresponding to
gluonic dissociation is due to
the diminishing overlap between the thermal gluon distribution and the gluonic dissociation cross section, it
is discussed in detail
in Ref.~\cite{nendzig}.  In contrast, collisional damping, derived from the imaginary part of the potential
which is directly
proportional to the temperature, leads to $\Gamma_{c,nl}$ increase monotonically with temperature.\\

\begin{figure}
\includegraphics[width=0.49\textwidth]{ratio.eps}
\caption{(Color online) The ratio between the decay widths of collisional damping and gluonic dissociation for
$J/\psi$ as
a function of temperature is shown. In the inset, the ratio of individual decay widths to the net decay width
is illustrated
with temperature.}
\label{ratio}
\end{figure}

In Fig.~\ref{ratio}, the ratio of the decay widths for collisional damping and gluonic dissociation is
presented as
a function of temperature. This analysis aims to investigate the impact of each mechanism on the dissociation
of $J/\psi$ in
an evolving medium. The fall in the ratio up to a temperature of approximately 250 MeV depicts that
$\Gamma_{gd,nl}$ increases
at $\rm T \lesssim 250$ MeV. Subsequently, it starts decreasing with increasing temperature, leading to a rise
in the ratio of
$\Gamma_{c,nl}$ to $\Gamma_{gd,nl}$ at $\rm T \gtrsim 250$ MeV. From this ratio, it can be seen that
$\Gamma_{c,nl}$  increases
with temperature. Moreover, to quantify the contributions of $\Gamma_{c,nl}$ and $\Gamma_{gd,nl}$ on $J/\psi$
dissociation,
their ratio with the total decay width $\Gamma_{D,nl}$ as a function of temperature is shown in the inset of
Fig.~\ref{ratio}.
In this inset, the ratio of $\Gamma_{c,nl}$ to $\Gamma_{D,nl}$ indicates that collisional damping is the
dominant dissociation
mechanism, as the ratio  varies around 0.9. In comparison, the ratio  $\frac{\Gamma_{c,nl}}{\Gamma_{gd,nl}}
\approx 0.1$,
illustrates the marginal effect of gluonic dissociation  on $J/\psi$ than the collisional damping.


\subsection{Time-dependent Schr$\ddot{o}$dinger equation: Nonadiabatic evolution of quantum states}

First, to study the nonadiabatic behavior of charmonium states, it is essential to ensure that the evolution
rate
exceeds the transition rate. We define the evolution timescale, \(\tau_{\text{ev}}\), as \(T \times
\frac{d\tau}{dT}\), while the transition timescale, \(\tau_{\text{tr}}\), represents the time associated with
transitions between
different energy states, specifically given by \(2\pi/\Delta E\) fm. Using this, we estimate
\(\tau_{\text{ev}}\) to be
approximately 0.3 fm during the thermalization phase, where the temperature evolution follows a Gubser-type
profile
with viscous correction. This estimation assumes an initial system size of $\sim$1.5 fm, which is relevant to
high-multiplicity \(p+p\) collisions. In contrast, the transition timescale \(\tau_{\text{tr}}\) for
charmonium states
is calculated to be around 4.0 fm.\\

Now, coming back to the charmonium evolution; the charmonia are hypothesized to form during the initial stages
of
collision. Utilizing a bottom-up thermalization approach rooted in QCD kinetic theory, it can be argued that
from an
initially interacting out-of-equilibrium state, a
thermalized medium is reached at a subsequent time marked as $\tau_{\text{Hydro}}$. Right after the collision
(well
before thermalization).  The evolution of the initial state of charmonia from $\tau = 0$ to
$\tau_{\text{Hydro}}$, can
be determined by solving the \textit{zero-temperature} Hamiltonian ~\cite{Wong:1995jf},

\begin{equation}\label{eq:H_0}
    H_0 = \frac{\vec{p}^2}{2M}  + \sigma r - \frac{4}{3} \frac{\alpha_s}{r}
\end{equation}

Here, $M$ denotes the reduced mass of the quark-antiquark system. However, as thermalization occurs in the
medium, the
zero-temperature Hamiltonian evolves into its finite temperature counterpart ~\cite{Karsch:1987pv},
\begin{equation}\label{eq:H_t}
    H(\tau)= \frac{\vec{p}^2}{2M} + Re(V)
\end{equation}
where $Re(V)$ is the real part of the potential given in Eq.~(\ref{pot}).\\

The time-dependent nature of
the Hamiltonian arises from the temporal variation of temperature. As the system expands further, the medium
temperature
eventually drops below the threshold for hadronization, causing the Hamiltonian to revert to a
zero-temperature state.
It has been contended in \cite{Bagchi:2023vfv} that the adiabatic approximation for the evolution of the
quantum bound
state of charmonia may not hold, as the  Hamiltonian evolves quite rapidly for medium produce in $p+p$
collision.\\

Due to the rapid evolution of the medium, the initial quarkonia states experience nonadiabatic evolution,
which may
cause transitions to states orthogonal to their initial configurations. Consequently, the survival probability
of the
initial state is affected. To determine this survival probability, we solve the time-dependent Schrödinger
equation,
which can be expressed as follows:\\

\begin{eqnarray} \label{eq:sc-eq}
    -\frac{1}{2M}\nabla^2 \psi + V(r)\psi = i\frac{\partial \psi}{\partial \tau}
\end{eqnarray}
In spherical polar coordinates $\nabla^2$ can be written as
\begin{equation}
\nabla^2 =\frac{1}{r^2}\left[\frac{\partial}{\partial r}\left(r^2\frac{\partial}{\partial r}\right) +
\frac{1}{\sin\theta}\frac{\partial}{\partial \theta}\left(\sin\theta\frac{\partial}{\partial \theta}\right) +
\frac{1}{\sin^2\theta}\frac{\partial^2}{\partial \phi^2}\right]
\end{equation}
and $V(r)= \frac{\sigma}{\mu}(1 - \exp(-\mu r)) - \frac{4}{3}\alpha_s \exp(-\mu r)/r$. For simplicity, we
are
considering potential is spherically symmetric, one can write $\psi=R(r)\psi(\theta,\phi)$ and  separate the
radial
part of Schr$\ddot{o}$dinger equation as
\begin{equation}\label{eq:radial_equn}
 \left[ -\frac{1}{2M} \frac{d^2}{dr^2} + V_{\text{eff}}(r) \right] u(r) = E u(r)
\end{equation}
where $V_{\text{eff}}(r) = \frac{\hbar^2 }{2 m r^2} l(l+1) + V(r)$,  $u(r) = rR(r)$ and $E$ is energy
eigenvalue,
represents the binding energy.\\

Here we solve the time-dependent Schrödinger equation [Eq.~(\ref{eq:radial_equn})] for the time-dependent
Hamiltonian
shown in Eq.~(\ref{eq:H_t}) using the Crank-Nicolson method (\cite{Crank_Nicolson_1947}) to obtain the survival
probability of a particular initial state. The initial states, charmonia bound states ($J/\psi$, $\psi$(2S),
and
$\chi_{c}$(1P)), evolve with time until the temperature drops below the QGP threshold temperature $T_c$. The
survival
probability of particular charmonium states can be calculated by taking the overlap integration of the final
wave
function with the initial zero-temperature charmonium state.\\

If we consider $|J/\psi\rangle$, $|\psi(2S)\rangle$, and $|\chi_{c}(1P)\rangle$ states to represent bound
states of the
initial zero-temperature Hamiltonian [Eq.~(\ref{eq:H_0})], and $\psi(\tau)$ represents the evolving wave
function, the
survival probability of $|J/\psi\rangle$, $|\psi(2S)\rangle$, and $|\chi_{c}(1P)\rangle$ at $\tau = \tau_c$
can be
represented as:
\begin{eqnarray}\label{eq:sur_prob}
 P_{J/\psi}  &=&|\langle\psi(\tau_c)| J/\psi\rangle|^2 \\
 P_{\psi(2S)}  &=&|\langle\psi(\tau_c)| \psi(2S)\rangle|^2 \\
 P_{\chi_{c}(1P)}  &=&|\langle\psi(\tau_c)|\chi_{c}(1P)\rangle|^2
\end{eqnarray}


\subsection{The regeneration factor}

In addition to the gluonic excitation of a color-neutral state to a color-octet state, gluonic deexcitation
from the
color-octet to the neutral state is also feasible. Consequently, charmonia gets regenerated in the QGP medium
through
this process. The regeneration is significant in heavy-ion collisions at LHC energies due to the abundant
production of
$c\bar{c}$ pairs in the hot QGP medium, which regenerate charmonia through recombination of $c\bar{c}$. While,
in
smaller systems like $p+p$ collisions, the production of $c\bar{c}$ pairs is relatively low, making the
regeneration
due to the coalescence less probable. However, regeneration due to gluonic deexcitation plays an important
role in
estimating charmonium production in such a small collision system (discussed in detail in
Ref.~\cite{Singh:2021evv}).
This deexcitation is calculated in terms of the regeneration cross section $\sigma_{f,nl}$ for charmonium by
employing
the detailed balance of the gluonic dissociation cross section $\sigma_{d,nl}$~\cite{Singh:2018wdt}:

\begin{equation}
 \sigma_{f,nl} = \frac{48}{36}\sigma_{d,nl} \frac{(s-M_{nl}^{2})^{2}}{s(s-4\;m_{c}^{2})}
\end{equation}
where $s$ is the Mandelstam variable, related with the center-of-mass energy of $c\bar c$ pair, given as; $s =
(p_c
+ p_{\bar{c}})^2$, where $ p_c$ and $ p_{\bar{c}}$ are four momenta of  $c$ and $\bar{c}$, respectively.\\

Finally, we have obtained the recombination factor $\Gamma_F$ by taking the thermal average of the product of
recombination cross section and relative velocity $v_{rel}$ between $c$ and $\bar{c}$:

\begin{equation}
\Gamma_{F,nl}=<\sigma_{f,nl}\;v_{rel}>_{p_{c}},
\end{equation}

\begin{equation}
v_{rel} =
\sqrt{\frac{({p_{c}^{\mu}\;
p_{\bar{c} \mu}})^{2}-m_{c}^{4}}{\textbf p_{c}^{2}\;\textbf p_{\bar{c}}^{2}
+ m_{c}^{2}(\textbf p_{c}^{2} + \textbf p_{\bar{c}}^{2}  + m_{c}^{2})}}
\end{equation}\\

Since the gluonic dissociation increases with the increase in temperature, it leads to the production of a
substantial
number of octet states in a high-multiplicity events. Such that the deexcitation of $c\bar c$ octet states to
$J/\psi$
enhances the regeneration of $J/\psi$ in high-multiplicity events conferred with relatively low multiplicities.

\subsection{The quantified yield}

Modification of the charmonium yield in the medium due to collisional damping, gluonic dissociation, and
regeneration
is
obtained by combining all these mechanisms in one transport
equation~\cite{thews1,thews2,Singh:2015eta,Singh:2018wdt,Singh:2021evv}:

\begin{equation}
 \frac{d N_{J/\psi(nl)}}{d\tau} = \Gamma_{F,nl} N_{c}~N_{\bar{c}}~[V(\tau)]^{-1} - \Gamma_{D,nl} N_{J/\psi(nl)}
\label{tq}
\end{equation}

The first term on the right-hand side of Eq.(~\ref{tq}) is a gain term, and the second is the loss term. Here,
$V(\tau)$
is the dynamic volume of the evolving medium. We assume that initially, the number of charms ($N_{c}$) and
anticharm
quarks $(N_{\bar{c}})$ are produced in equal numbers,
$N_{c}$ = $N_{\bar{c}}$ = $N_{c\bar{c}}$. Equation~(\ref{tq}) can be  solved analytically under the
assumption of
$N_{J/\psi}(nl) < N_{c\bar{c}}$ at $\tau_{0}$~\cite{Singh:2015eta,Singh:2018wdt,Singh:2021evv}:

\begin{align}
N_{J/\psi(nl)}^{f}(\tau_{QGP},p_{T})\; = \; \epsilon_1(\tau_{QGP},p_{T}) \bigg[
N_{J/\psi(nl)}^{i}\;\; \nonumber
\\ + N_{c\bar{c}}^{2} \int_{\tau_{0}}^{\tau_{QGP}} \Gamma_{F,nl}(\tau,p_{T})
[V(\tau)\epsilon_2(\tau,p_{T})]^{-1} d\tau \bigg]
\label{tq1}
\end{align}

Here, $N_{J/\psi(nl)}^{f}(\tau_{QGP}, p_{T})$ is the net number of charmonium formed during the QGP evolution
period
$\tau_{QGP}$. The quantities $N_{J/\psi(nl)}^{i}$ and $N_{c\bar{c}}$ represent the number of $J/\psi$ and
$c\bar{c}$ pairs formed during the initial hard scattering, respectively. Input for  $N_{J/\psi(nl)}^{i}$ and
$N_{c\bar{c}}$ are taken from Ref.~\cite{Singh:2021evv} corresponding to $p+p$ collisions at $\sqrt{s}$ = 13
TeV.\\

In Eq.~(\ref{tq1}), $\epsilon_1(\tau_{QGP})$ and $\epsilon_2(\tau)$ are decay factors for the meson due to
gluonic
dissociation and collisional damping in the QGP with a lifetime of $\tau_{QGP}$, and $\tau$ represents the
evolution
time. These factors are calculated using the following expressions:

\begin{equation}
\epsilon_1(\tau_{QGP},p_{T}) = \exp{\left[-\int_{\tau_{nl}^{'}}^{\tau_{QGP}}
\Gamma_{D,nl} (\tau, p_{T}) d\tau\right]},
\end{equation}
and
\begin{equation}
\epsilon_2(\tau, p_{T}) = \exp{\left[-\int_{\tau_{nl}^{'}}^{\tau} \Gamma_{D,nl}(\tau^{\prime}, p_{T})
d\tau^{\prime}\right]}.
\end{equation}

After obtaining the $N_{J/\psi(nl)}^{f} (\tau_{QGP}, p_{T})$ from Eq.~(\ref{tq1}) we took the ratio to the
initially
produced charmonium, $N_{J/\psi(nl)}^{i}$ to quantifying the medium effect and called it survival probability
$S_{P}$. The survival probability of charmonium, due to gluonic dissociation and collisional damping along
with the regeneration effect is defined as:

\begin{equation}
S_{cgr}^{J/\psi}(p_{T}, mc) = \frac{N_{J/\psi(nl)}^{f}(p_{T},
mc)}{N_{J/\psi(nl)}^{i}(mc)}
\label{sp2}
\end{equation}
where ``$mc$'' stands for Multiplicity Class defined as $<dN_{ch}/d\eta>$.\\

It is assumed here that from the initial collision to the QGP endpoint  (at $\tau=0$ to $\tau = \tau_{qgp}$),
the
nonadiabatic evolution of charmonia states is a completely independent process with the other suppression
mechanisms
in QGP.  The net yield in terms of survival probability,  $S_{P}(p_{T},mc)$ is expressed as:\\
\begin{equation}
S_{P}^{J/\psi}(p_{T},mc) = S_{gc}^{J/\psi}(p_{T}, mc)\;P_{J/\psi}(p_{T}, mc).
\label{spf}
\end{equation}
Further, we incorporate the feed-down correction, which refers to the decay of higher excited quarkonium states
into lower ones, such as the decay of $\chi_c$ and $\psi(2S)$ into $J/\psi$. These higher excited states are
more susceptible to dissociating in the QGP due to their relatively small binding energies, leading to
sequential suppression. Consequently, the suppression of these excited states leads to a reduced feed-down
contribution to the $J/\psi$  yields. Therefore, while feed-down does not alter the intrinsic survival
probability of the $J/\psi$ itself, it affects  the inclusive  $J/\psi$ survival probability. Now, to determine
the feed-down correction in the inclusive  $J/\psi$ survival probability, we calculate the ratio between the
net initial and final numbers of $J/\psi$. The net initial
number is derived by accounting for the feed-down from higher resonances into $J/\psi$ in the absence of
the QGP medium. This is expressed as $N_{J/\psi}^{in} = \sum_{J\ge I} C_{IJ} N(J)$, where $C_{IJ}$ represents
the branching
ratio for the decay of state $J$ into state $I$. The net final number of $J/\psi$ incorporates medium effects,
represented by the survival probability ($S_{P}(p_{T},b)$), along with feed-down: $N_{J/\psi}^{fi} =
\sum_{J\ge I}
C_{IJ} N(J) S_{P}(J)$. The overall generalized survival probability, including the feed-down correction, is
given
as~\cite{Singh:2018wdt}:

\begin{equation}
S_{P}(I) = \frac{\sum_{J\ge I} C_{IJ} N(J) S_{P}(J)}{\sum_{J\ge I} C_{IJ} N(J)}.
\end{equation}

\section{Results and Discussions}
\label{results}

In our investigation of charmonium yield modification ultrarelativistic proton-proton ($p+p$) collisions at
$\sqrt{s}
= $13 TeV under the above-mentioned circumstances, we have obtained the survival probability ($\rm S_{P}$),
the double
ratio (used for the direct comparison of two probabilities), and the variation of particle ratios with respect
to both
charged particle multiplicity ($<dN_{ch}/d\eta>$) and transverse momentum ($p_{T}$) at the midrapidity. This
study
analyzes a spectrum of suppression mechanisms alongside the regeneration process. We have methodically
categorized
these mechanisms into two distinct groups for clarity and detailed analysis: CGR and NAb. The ``CGR" group
encapsulates
mechanisms such as collisional damping, gluonic dissociation, and recombination processes, highlighting the
interactions that directly involve gluonic exchanges. On the other hand, ``NAb" focuses on the nonadiabatic
evolution
of charmonium states, considering the temporal evolution under the scenario when the reaction time is so short
that the
transition amplitude is described as the overlap of these states. The combined effects, both CGR and NAb, are
presented
in a dataset labeled ``Net", showcasing the intertwined relationship and net impact of these complex
mechanisms on the
charmonium yield in such high-energy collisions.


\subsection{Multiplicity-dependent yield}

\begin{figure}[htp]
\includegraphics[width=0.49\textwidth]{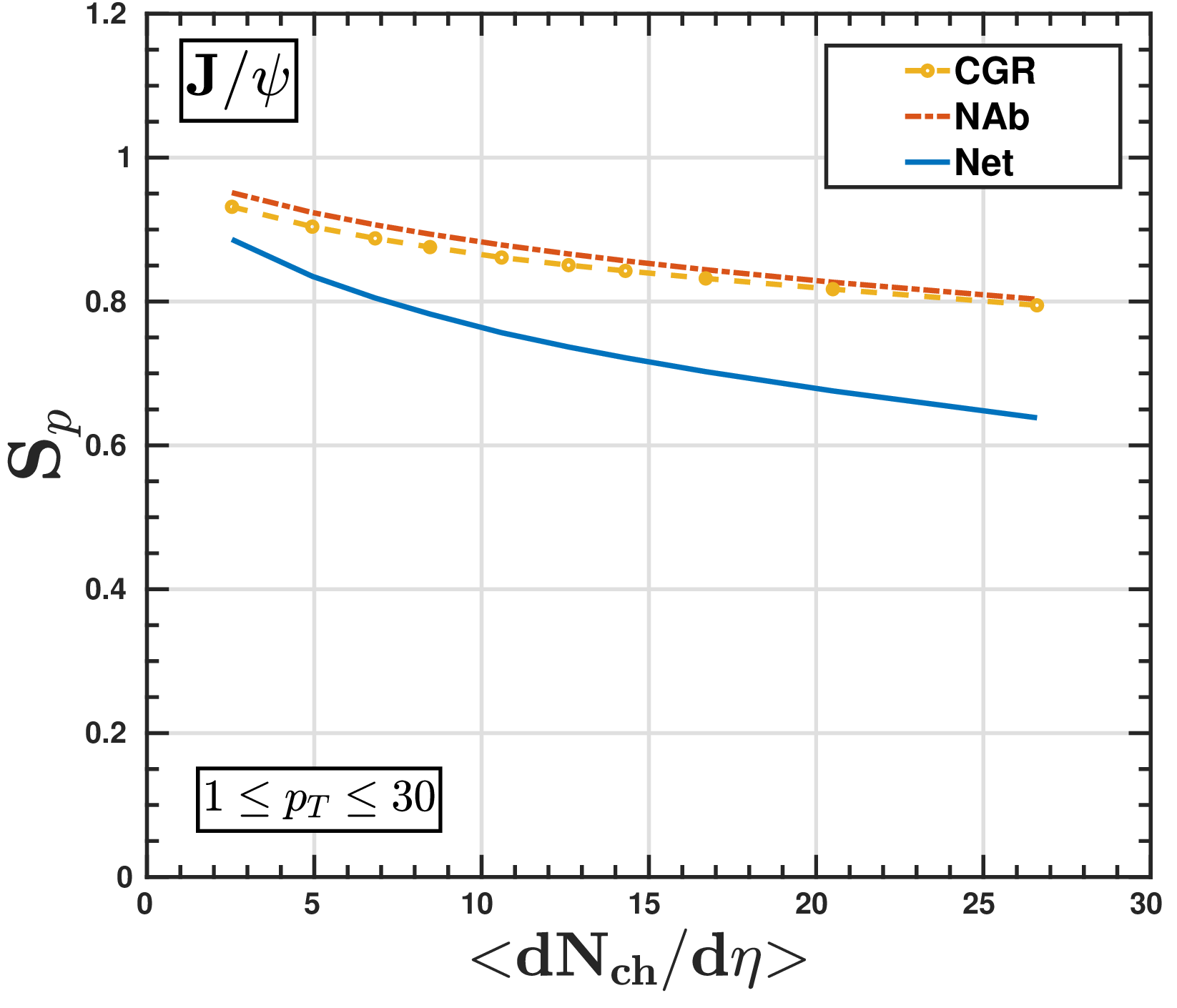}
\includegraphics[width=0.49\textwidth]{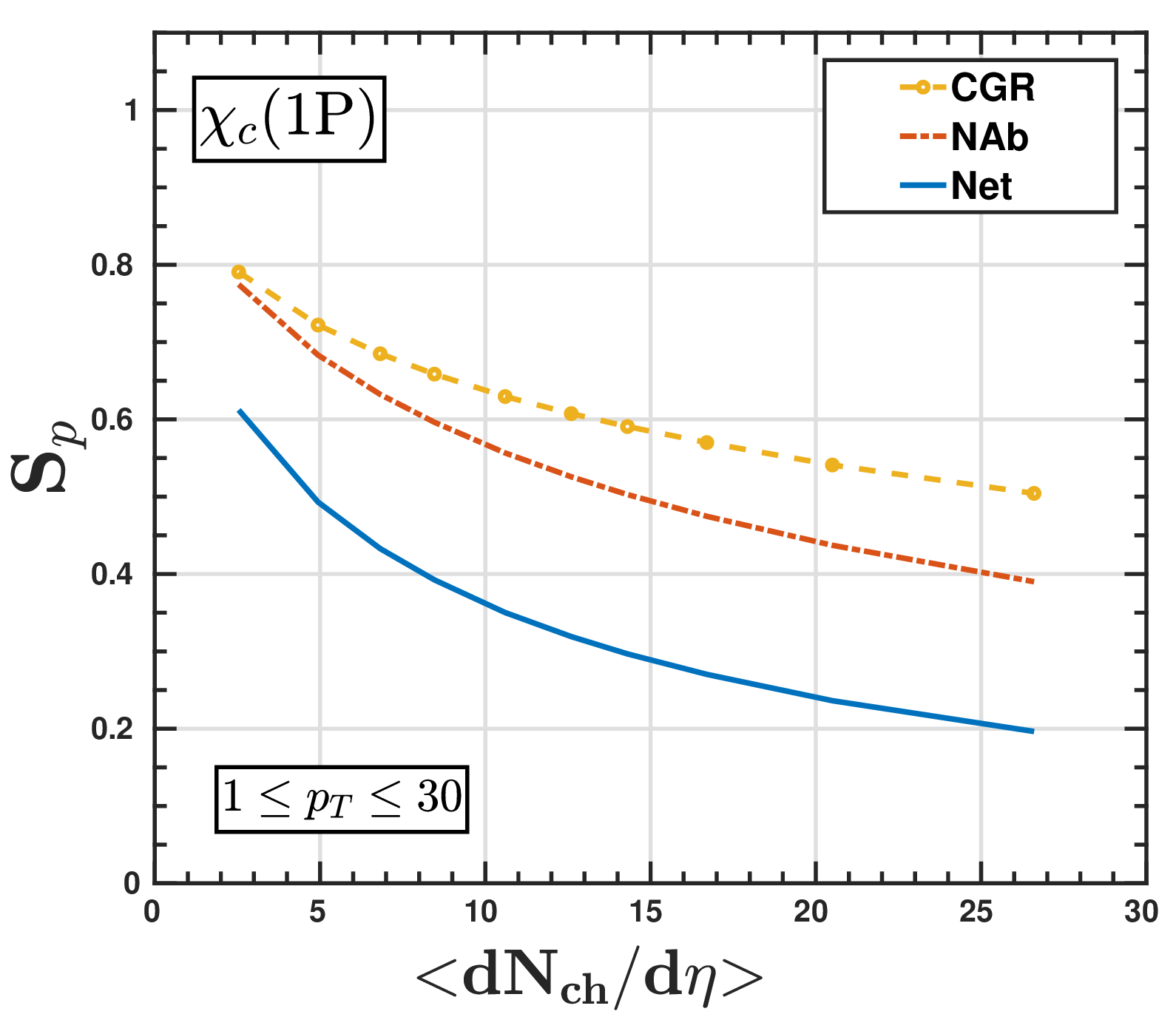}
\includegraphics[width=0.49\textwidth]{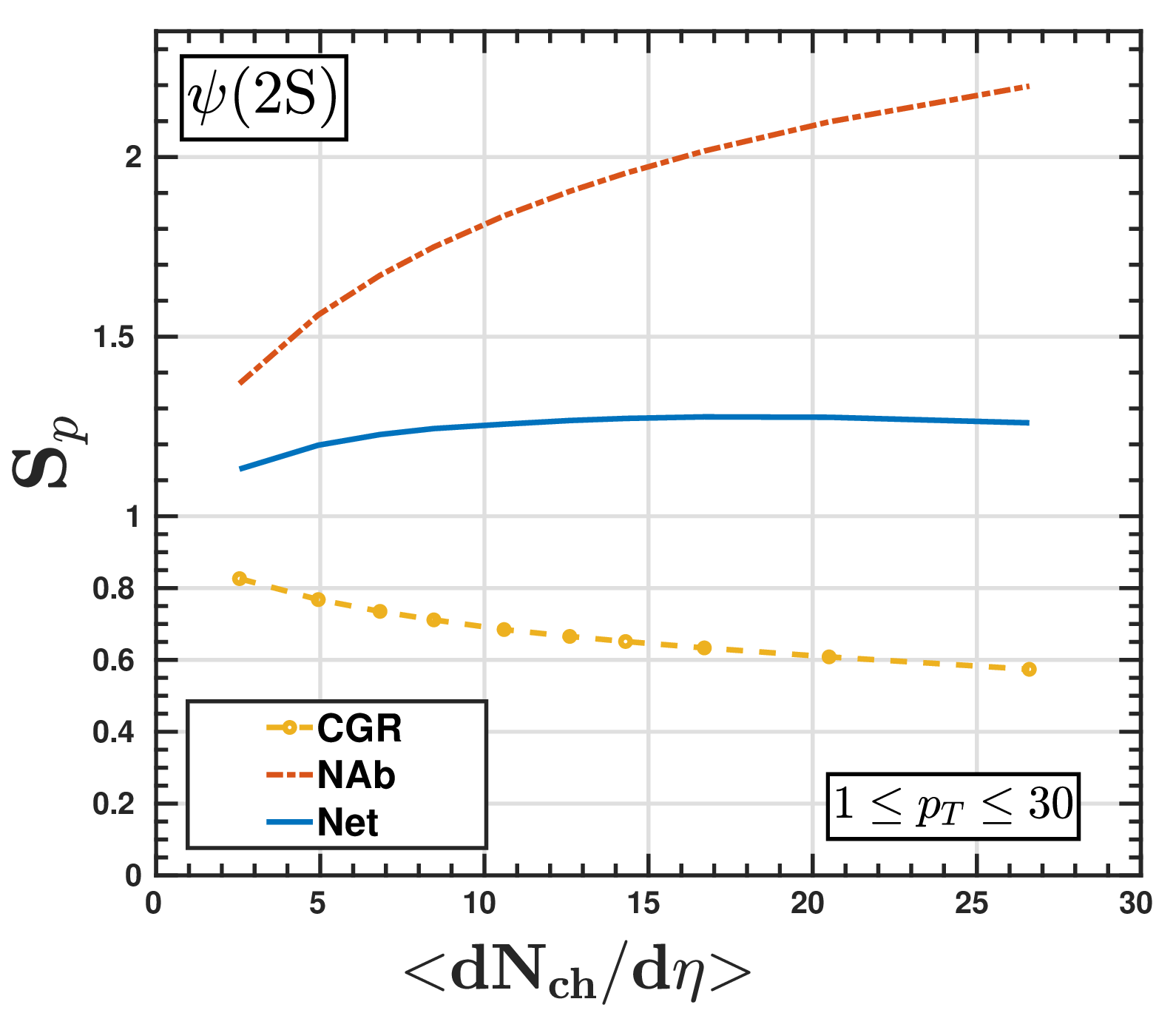}
\caption{(Color online) Survival probability $S_{P}$ as a function of multiplicity is shown for $J/\psi$,
$\chi_{c}$(1P)
and $\psi$(2S) at midrapidity corresponding to  $p+p$ collision at $\sqrt{s}=$13 TeV.}
\label{fig:2}
\end{figure}

The $p_{T}$-integrated charmonium yield modification in terms of $\rm S_{P}$ with event
multiplicity at mid rapidity has been explored using the charmonium distribution function $1/E_{T}^{4}$
as discussed in the Ref.~\cite{Singh:2018wdt, Singh:2021evv}. The yield modification of $J/\psi$,
$\chi_c$ and $\psi$(2S) shown in Fig.~\ref{fig:2}, predicts that suppression due to CGR is relatively large for
$\chi_{c}$(1P) than $J/\psi$ and $\psi$(2S), and it further increases with increasing multiplicity. The CGR
and NAb
independently predict about 20\% suppression for $J/\psi$ at highest multiplicity, as shown in
Fig.~\ref{fig:2}. However, $J/\psi$
suppression due to NAb is slightly less than CGR at low-multiplicity events. Conversely, NAb predicts
substantial suppression for
$\chi_{c}$(1P) compared to CGR. In contrast to $J/\psi$ and $\chi_c$, $\psi$(2S)
experiences an enhancement due to the NAb approach, increasing with multiplicity. Normally, the average radius
of
$\chi_{c}$(1P) is larger than that of $J/\psi$, and the average radius of $\psi$(2S) is even larger than both.
As the
average radius increases, dissociation due to nonadiabatic evolution also increases. Consequently,
$\chi_{c}$(1P)
undergoes more dissociation compared to $J/\psi$, and $\psi$(2S) should, in principle, experience even greater
dissociation than both $J/\psi$ and $\chi_{c}$(1P). Despite this, $\psi$(2S) exhibits an enhancement due to
nonadiabatic transitions from $J/\psi$ to $\psi$(2S), as described in Ref.~\cite{Bagchi:2024gud}.  The
combined effects
of CGR and NAb, represented through ``Net'',  lead to up to 40\% suppression for $J/\psi$ and 80\% for
$\chi_{c}$(1P) with increasing
multiplicity. For $\psi$(2S), the combined effects reduce the enhancement to some extent but are unable to
transform it
into suppression.\\

\begin{figure}
\includegraphics[width=0.5\textwidth]{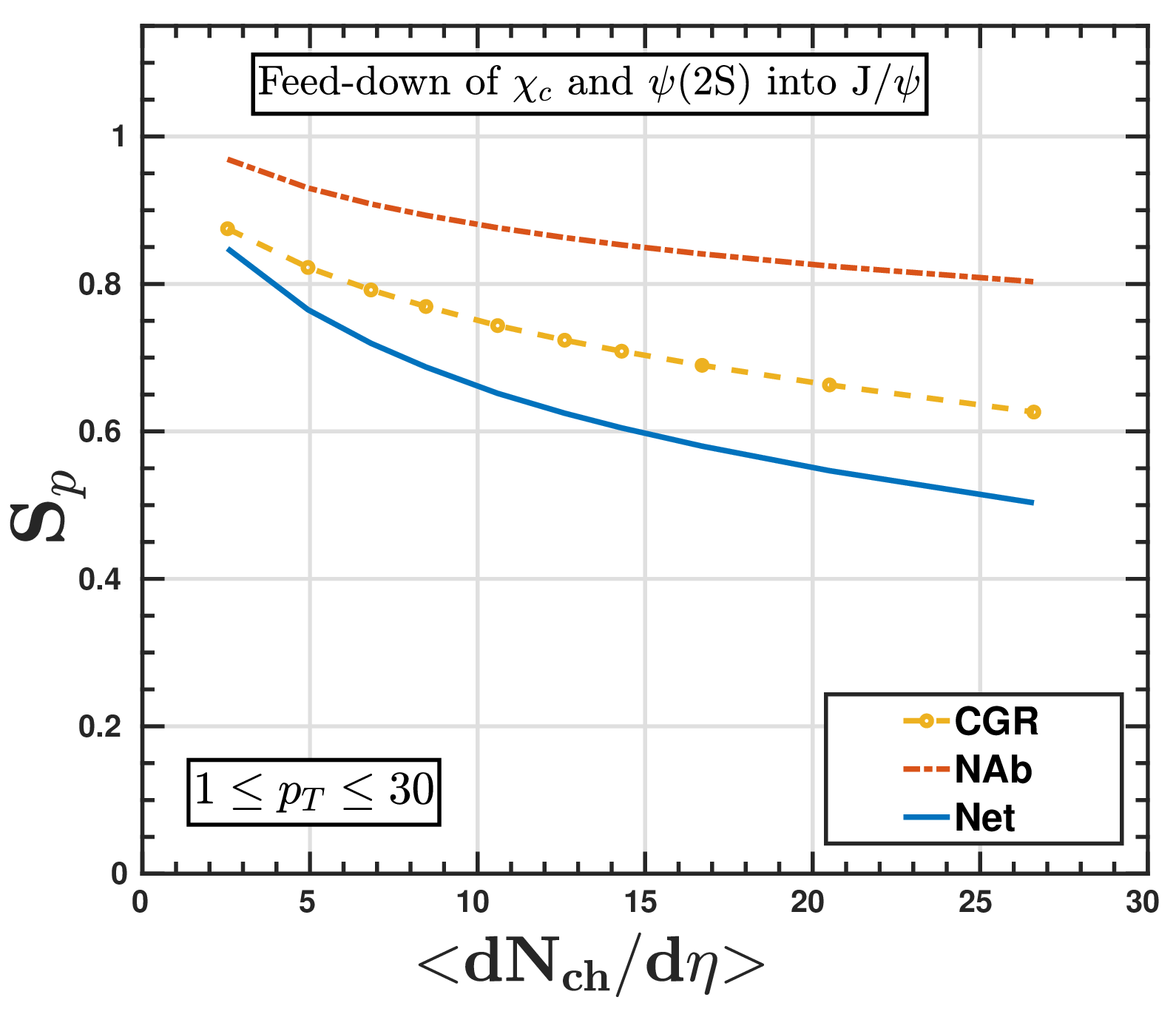}
\caption{(Color online) Survival probability $S_{P}$ as a function of multiplicity is shown for $J/\psi$,
considering
the feed-down of $\chi_{c}$(1P) and $\psi$(2S) into $J/\psi$ at midrapidity corresponding to  $p+p$ collision
at
$\sqrt{s}=$13 TeV.}
\label{fig:3}
\end{figure}

In Fig.~\ref{fig:3}, the feed-down of the  $\chi_{c}$(1P) and $\psi$(2S) into $J/\psi$
further increases the suppression for $J/\psi$ at all the multiplicity classes. The feed-down corresponding to
the NAb
only predicts maximum suppression up to 20\% for $J/\psi$, which is almost the same as its prediction in the
absence of
the feed-down correction. While considering feed-down correction, only incorporating the CGR process increases
the
suppression up to 40\% while twice as its earlier prediction (CGR without feed-down correction shown in
Fig.~\ref{fig:2}). Finally, the combined effects of CGR and NAb, with feed-down of higher resonances, lead to
50\% suppression for $J/\psi$ at high-multiplicity events in ultrarelativistic $p+p$ collisions at $\sqrt{s}$
= 13 TeV.\\

\begin{figure}[htp]
\includegraphics[width=0.49\textwidth]{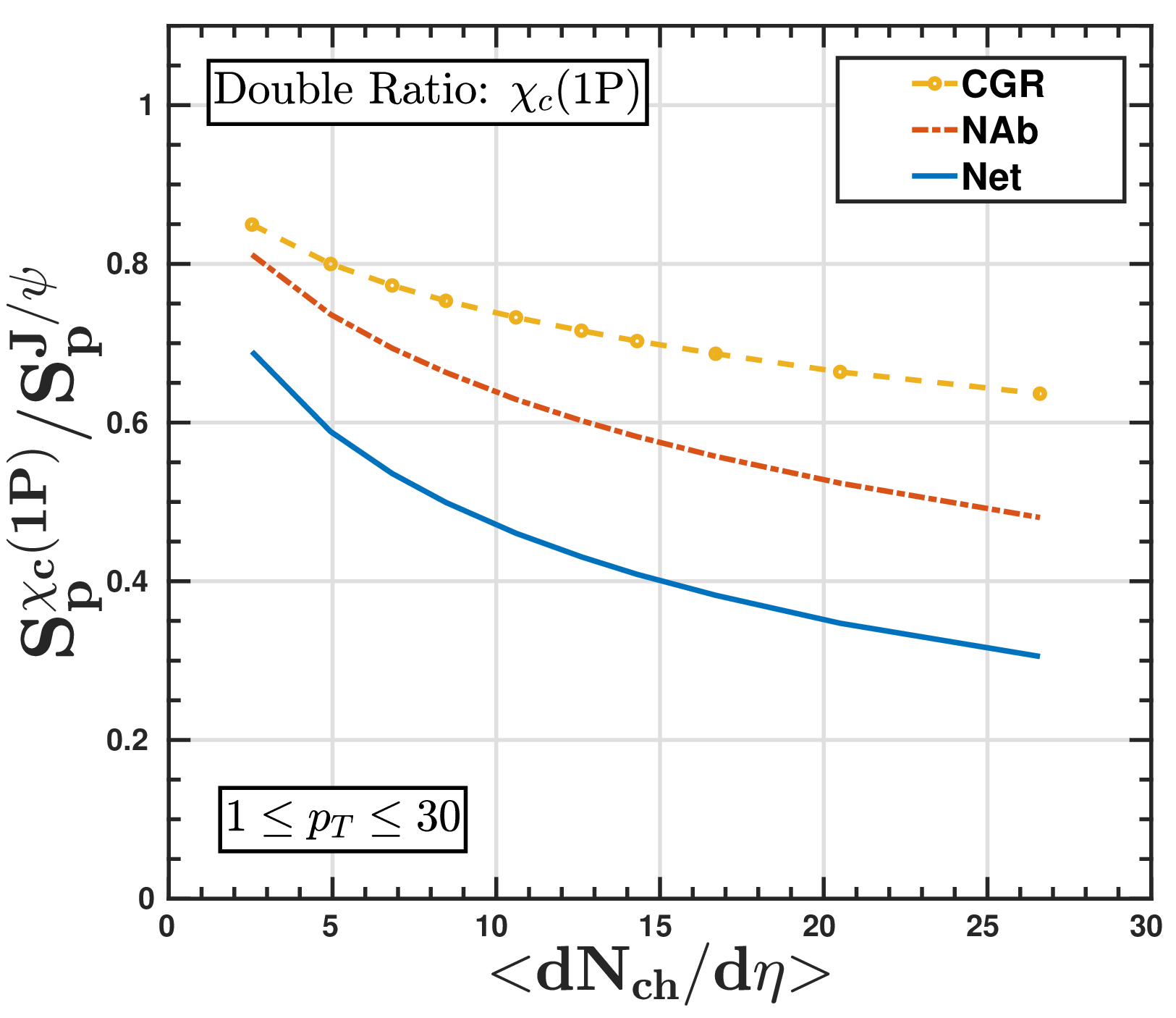}
\includegraphics[width=0.49\textwidth]{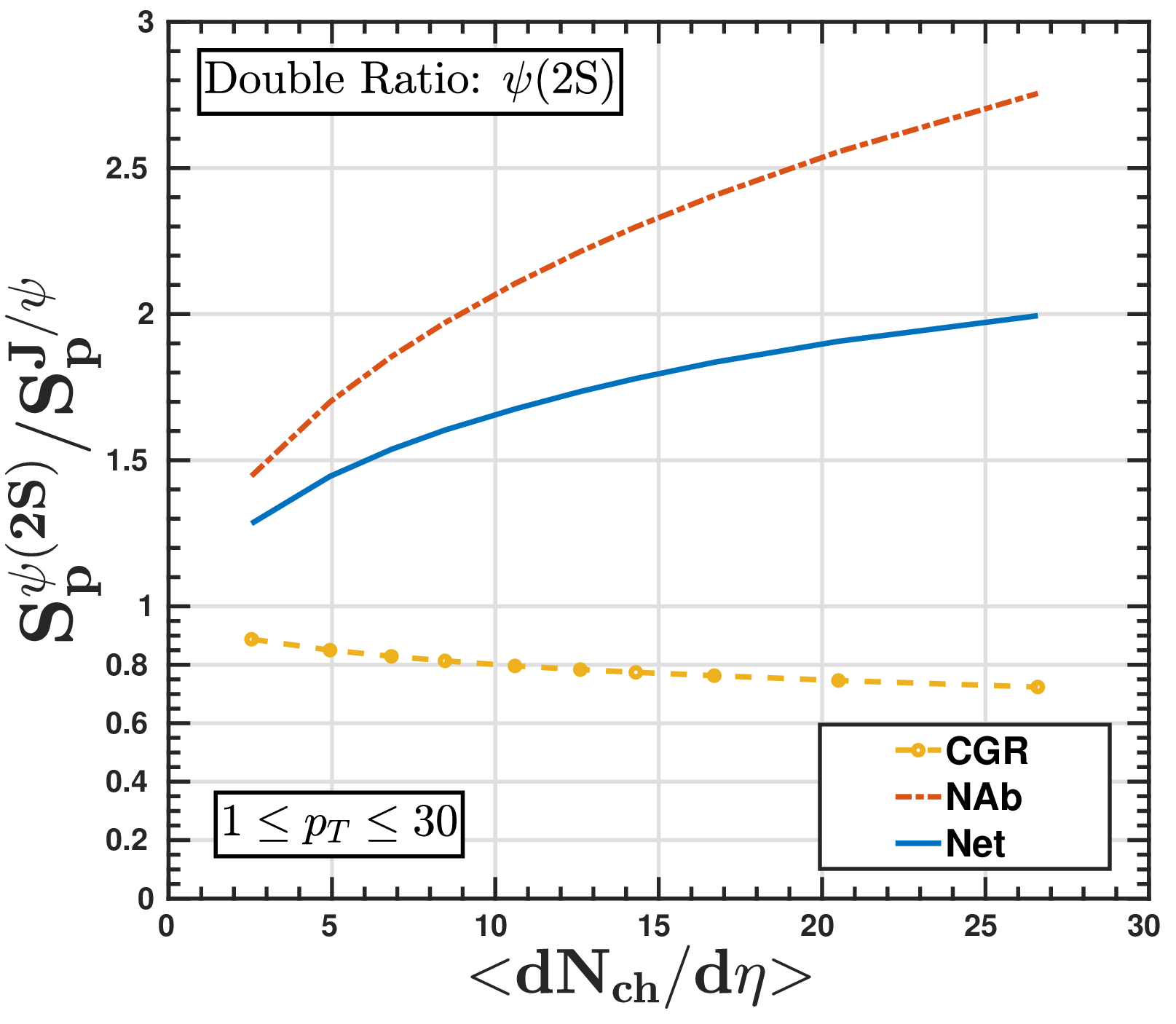}
\caption{(Color online) Double ratio as a function of multiplicity is shown for $\frac{\chi_{c}(1P)}{J/\psi}$,
 and
$\frac{\psi(2S)}{J/\psi}$,  at midrapidity corresponding to  $p+p$ collision at $\sqrt{s}=$13 TeV.}
\label{fig:4}
\end{figure}

In Fig.~\ref{fig:4} the survival probability ratios or double ratio between $\chi_{c}$(1P) to $J/\psi$ and
$\psi$(2S)
to $J/\psi$ are shown to quantify the relative yield modification of $\chi_{c}$(1P) and $\psi$(2S)
with respect to $J/\psi$. Experimental observations employ double ratios to ascertain that the medium, which
may have
existed in ultrarelativistic collisions whether, affects the $\psi$(2S) and $J/\psi$ yields differentially or
the
same. Notably, due to the technical difficulties in the observation of $\chi_{c}$(1P), the yield modification
and double
ratio for $\chi_{c}$(1P) with $J/\psi$ has not been reported in any of the ultrarelativistic heavy-ion
collision
experiments. However, the present study explored the $J/\psi$, $\chi_{c}$(1P) and $\psi$(2S) dynamics in the
medium and the impact
on their yield imposed by the medium. Figure~\ref{fig:4} depeicts that $\chi_{c}$(1P) experiences significant
suppression
compared to $J/\psi$ at high-multiplicity, whereas the suppression magnitude for $\chi_{c}$(1P) to $J/\psi$ is
relatively small at low-multiplicity. On the other hand,  the yield of $\psi$(2S) is considerably enhanced
compared to
$J/\psi$ due to the NAb mechanism, leading to a populated transition to $\psi$(2S) in the final state, which
increases
with multiplicity. However, CGR predicts that $\psi$(2S) is more suppressed than $J/\psi$ but less than
$\chi_c$ at the high-multiplicity
classes. While at low-multiplicity $\psi$(2S) and  $\chi_{c}$(1P) are almost equally suppressed. The
cumulative influence of CGR
and NAb predicts a multiplicity-dependent 30\% to 70\% suppression for $\chi_c$ compared to $J/\psi$ and
similarly estimates
an enhancement of approximately 130\% to 200\% for $\psi$(2S) relative to $J/\psi$.\\

\begin{figure*}[htp]
\includegraphics[width=0.49\textwidth]{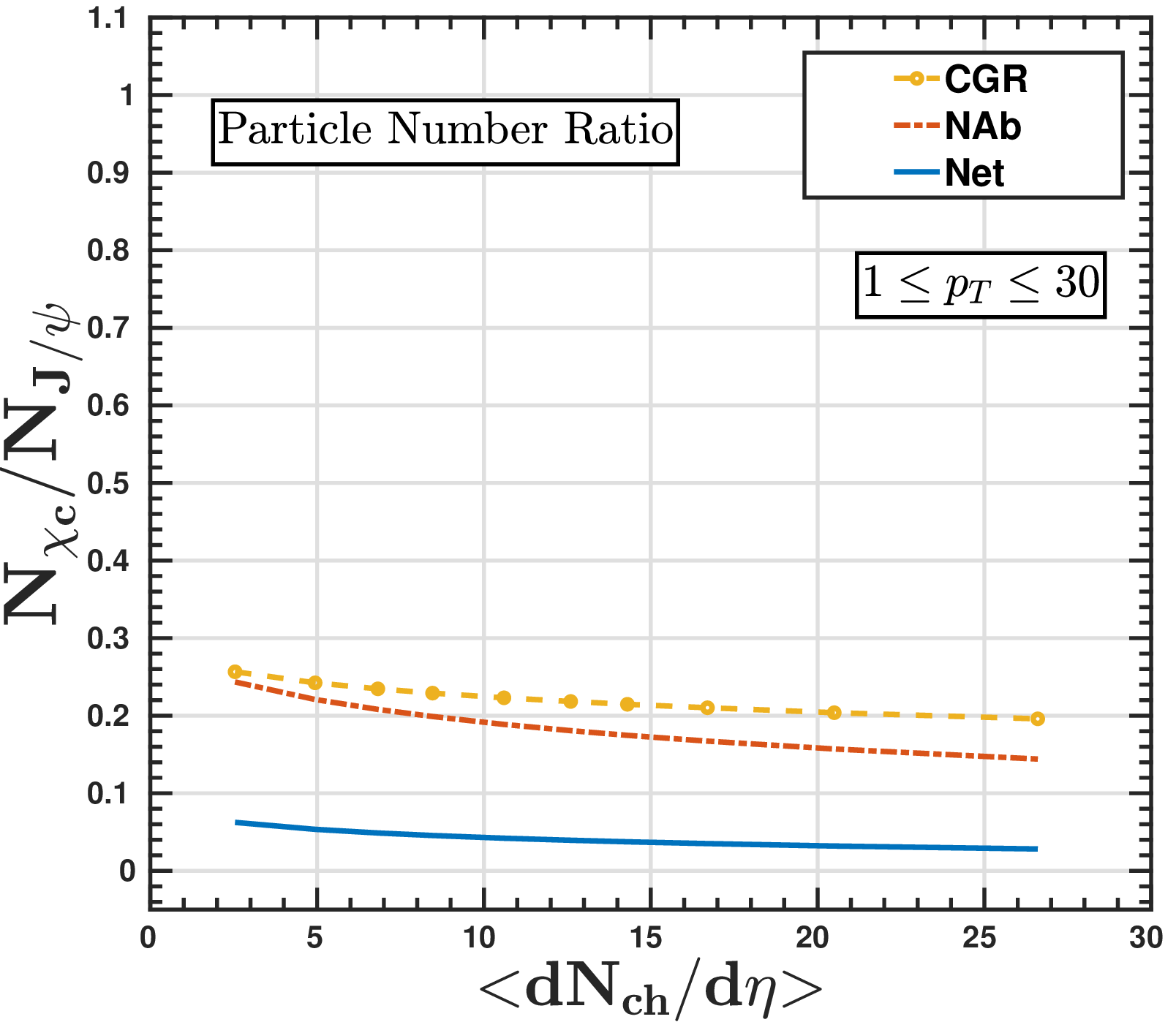}
\includegraphics[width=0.49\textwidth]{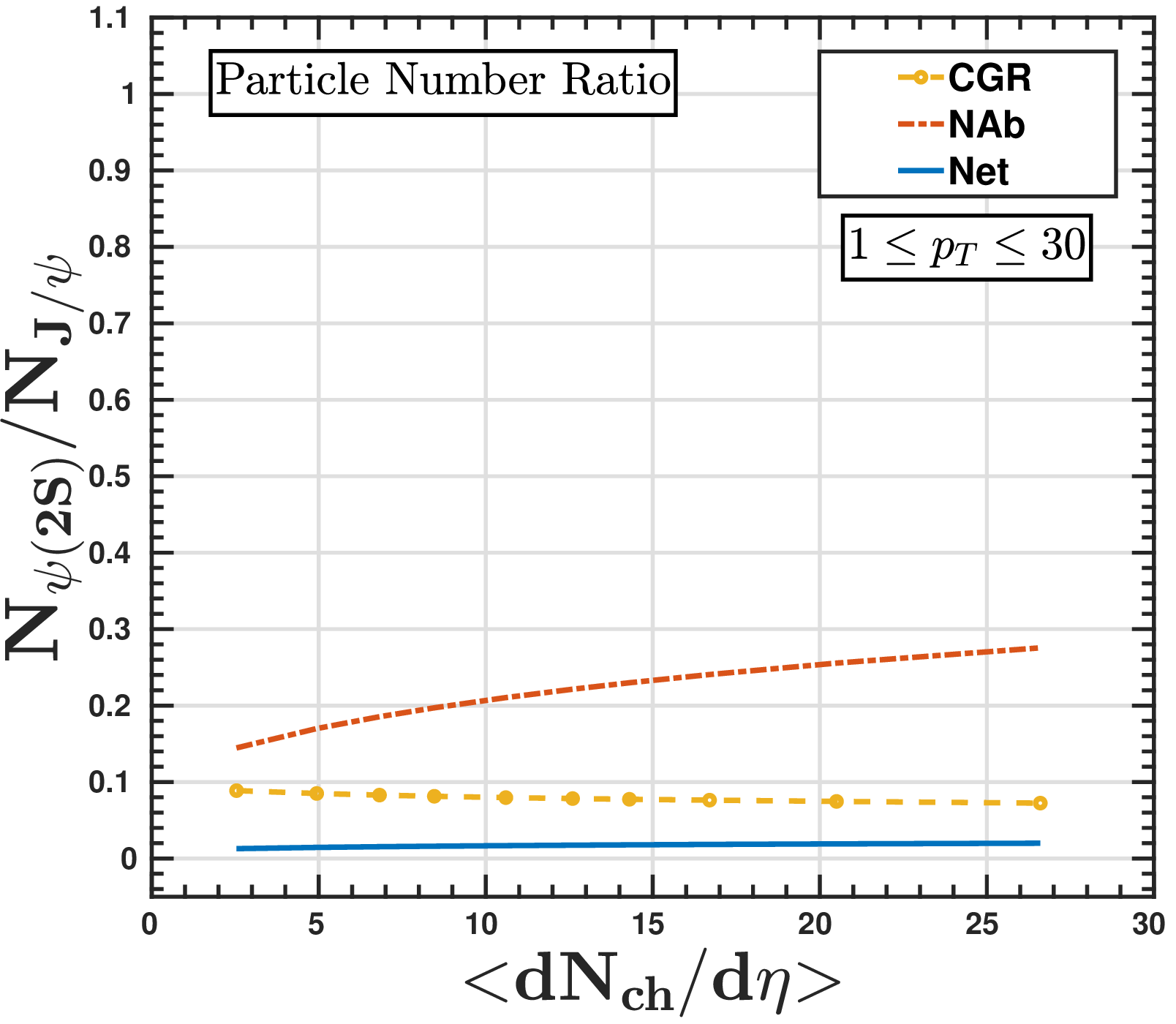}
\caption{(Color online) Particle number ratio  as a function of multiplicity is shown for
$\frac{\chi_{c}(1P)}{J/\psi}$,
 and $\frac{\psi(2S)}{J/\psi}$ at midrapidity corresponding to  $p+p$ collision at $\sqrt{s}=$13 TeV.}
\label{fig:5}
\end{figure*}

Furthermore, we extend the examination of survival probability ratios of the charmonium states in
ultrarelativistic
collisions to quantitatively assess the final numbers of $J/\psi$, $\chi_{c}$(1P), and $\psi$(2S) at the
chemical
freezeout boundary, as depicted in Fig.~\ref{fig:5}. These results align with previous observations,
indicating a
notably higher production and relatively less suppression of $J/\psi$ compared to $\chi_{c}$(1P) and
$\psi$(2S) during
the transportation from initial production to the QGP endpoint, i.e., $\rm T = T_{c}$. Additionally,
Fig.~\ref{fig:5}
suggests that $\chi_{c}$(1P) dissociation due to the CGR  mechanism is relatively smaller in comparison with
$\psi$(2S).
While the NAb mechanism effectively reduces the $\chi_{c}$(1P) and comparably predicts a large production for
$\psi$(2S). The $\chi_c$ yield decreases with increasing multiplicity for both CGR and NAb processes. The
$\psi$(2S) yield is almost steady at all the multiplicities corresponding to CGR while it increases with
multiplicity.
Meanwhile, the combined effects of CGR and NAb predict the survived production for $\chi_{c}$(1P) around 8\%
to 2\%
and approximately 0.5\% to 1\% for $\psi$(2S) with respect to $J/\psi$ depending on the multiplicity classes.


\subsection{Yield modification with transverse momentum}

The production of charmonia as a function of transverse momentum ($p_{T}$) provides valuable insights into the
physics
at both low and high $p_{T}$. We have also examined the impact of system size or, in this case, multiplicity
classes on
the charmonium yield over the considered $p_{T} $range. The chosen multiplicity classes include the lowest
multiplicity
class (Multi. Class X: 70 - 100\%), the highest multiplicity class (Multi. Class I: 0 - 1\%), and minimum
bias (Min. Bias: 0 - 100\%). To this end, we have computed the survival probability ($S_{P}$) as a function of
$p_{T}$
by averaging over the range of the corresponding multiplicity bins. The expression for the weighted average of
$S_{P}$
is given by~\cite{Singh:2018wdt}:

\begin{equation}
 S_{P}(p_{T}) = \frac{\sum_{i} S_{P}(p_{T},\langle b_{i} \rangle) W_{i}}{\sum_{i} W_{i}}
 \label{mib}
\end{equation}

The index $i = 1, 2, 3, ...$ represents the multiplicity bins. The weight function $W_{i}$ is defined as
$W_{i} =
\int_{b_{i\;min}}^{b_{i\;max}} N_{coll}(b)\pi\; b\; db$. The number of binary collisions $N_{coll}$ is
determined using
a Glauber model for $p+p$ collisions, which incorporates an anisotropic and inhomogeneous proton density
profile to
calculate $N_{coll}$~\cite{suman}. Also, we have obtained the impact parameter, $b$, for
$p+p$ collisions corresponding to multiplicity bins at $\sqrt{s}$ = 13 TeV using the above-mentioned Glauber
model.\\

\begin{figure}[htp]
\includegraphics[width=0.49\textwidth]{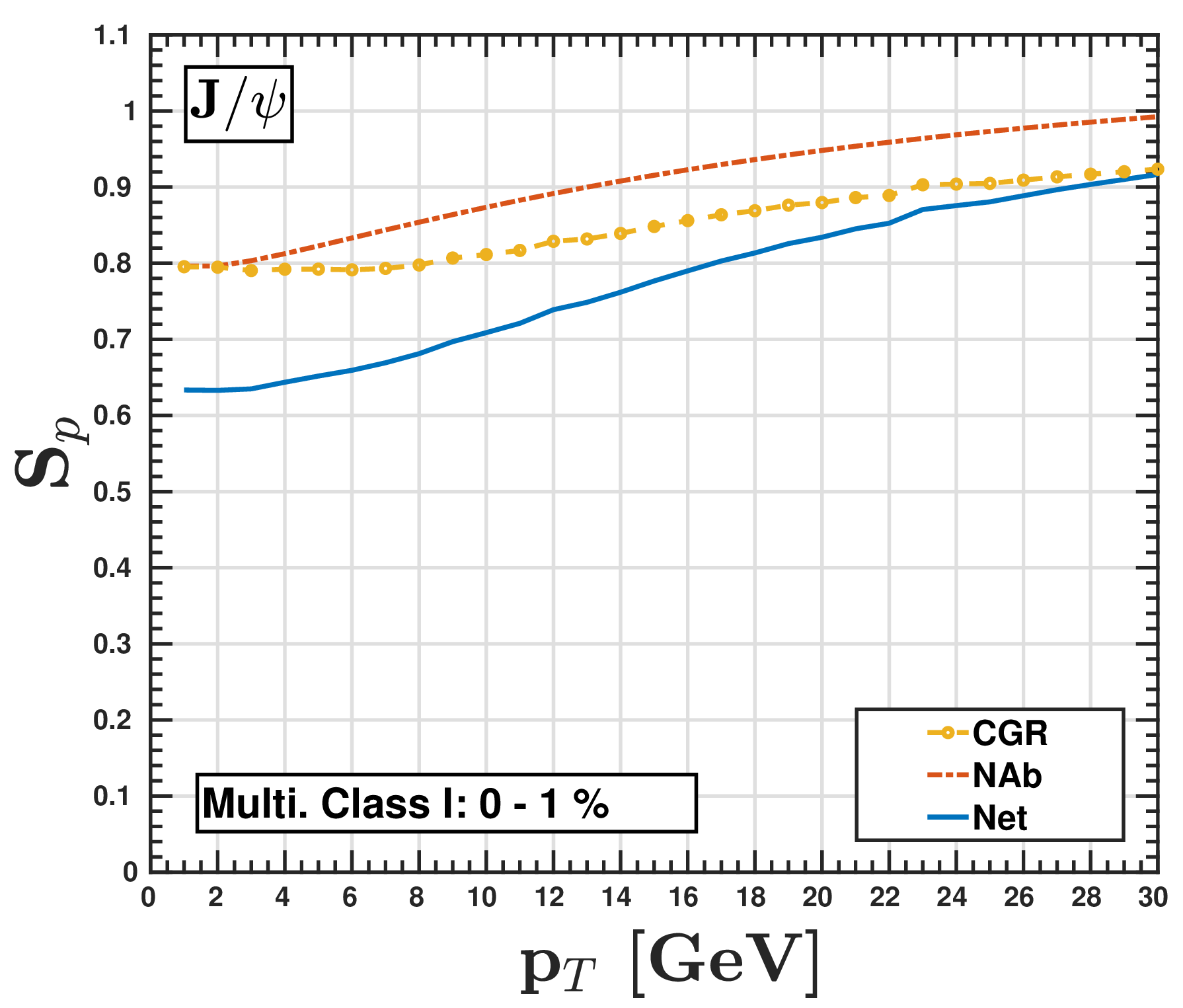}
\includegraphics[width=0.49\textwidth]{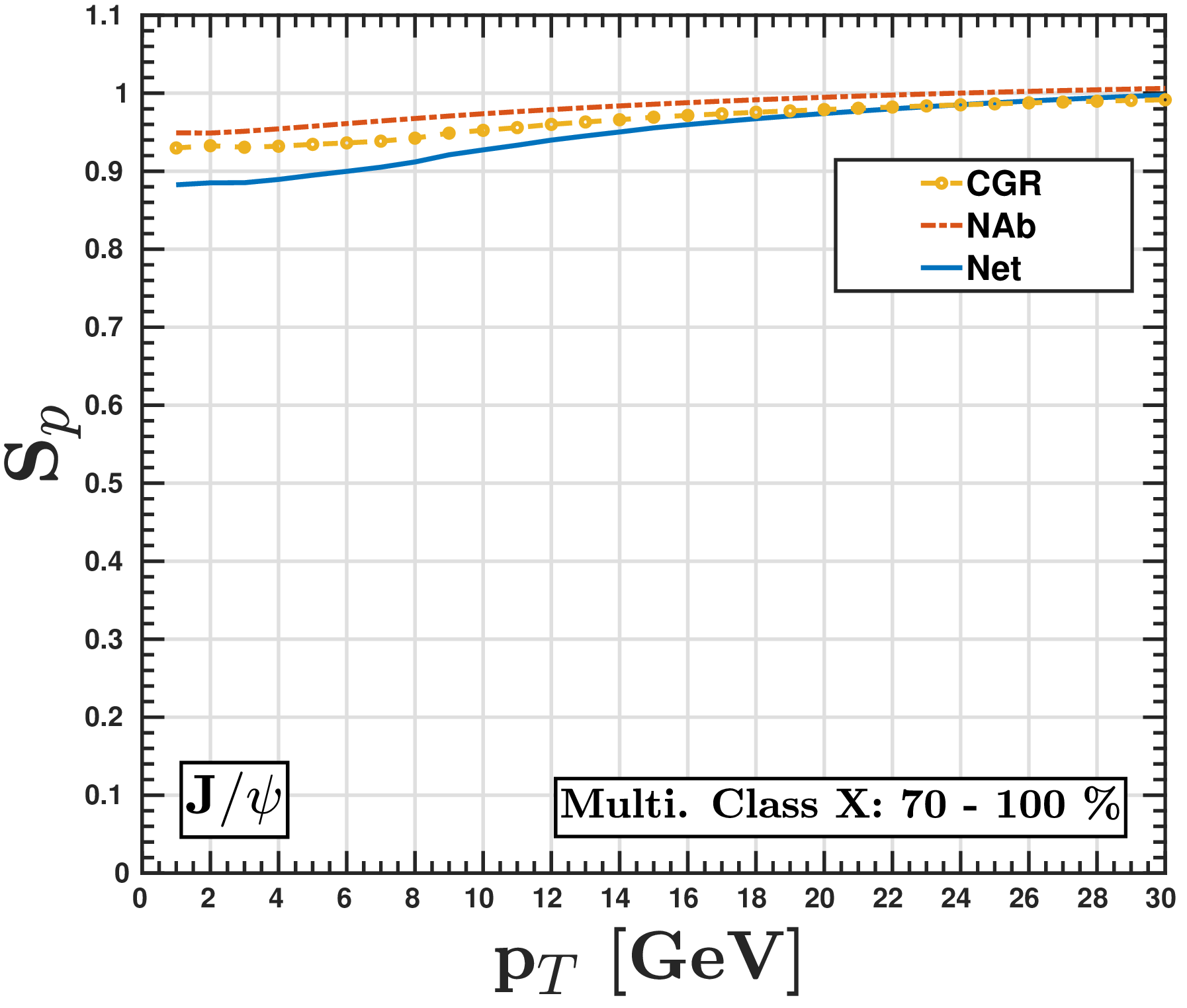}
\includegraphics[width=0.49\textwidth]{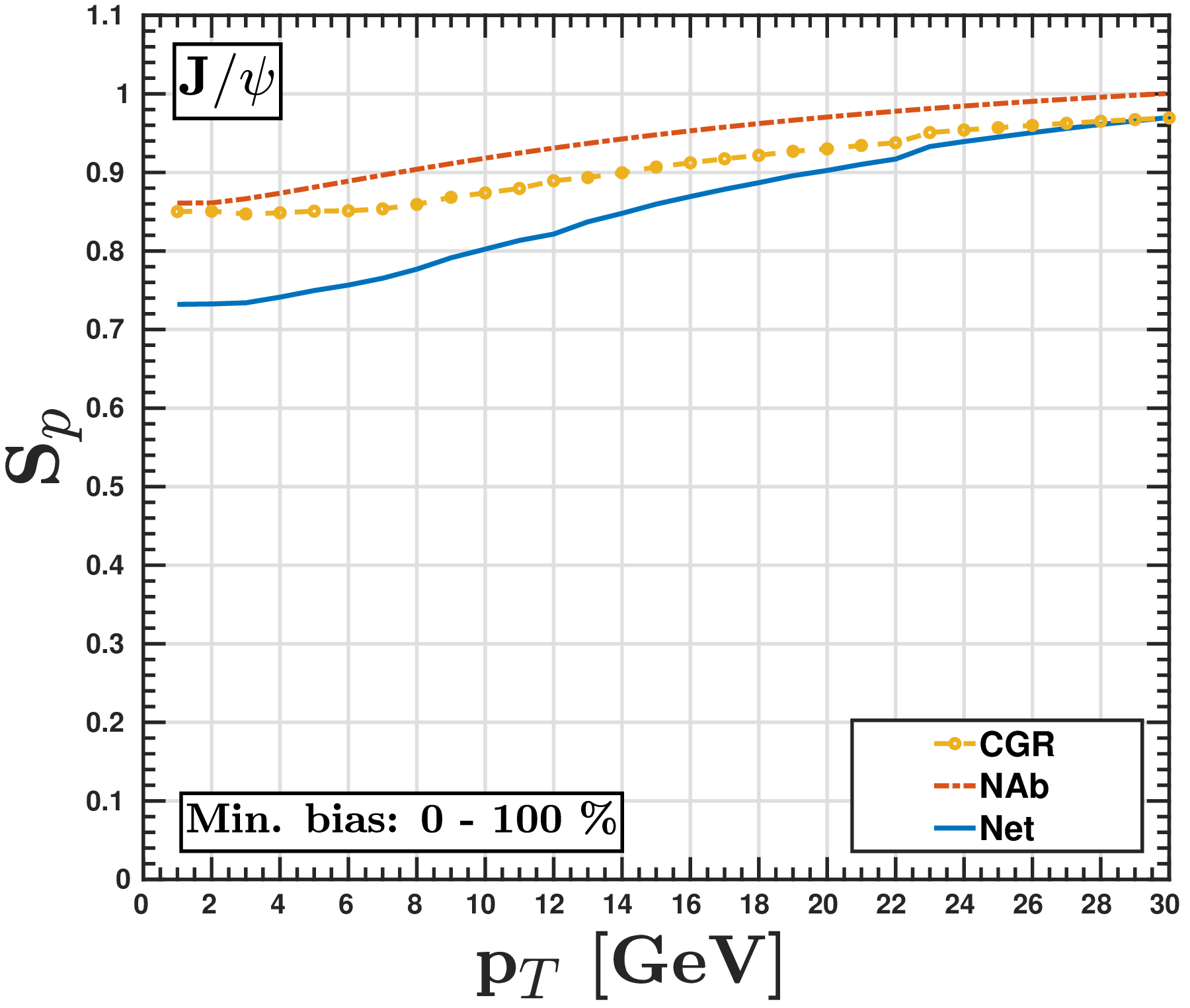}
\caption{(Color online) Survival probability $S_{P}$ as a function of $p_{T}$ is shown for $J/\psi$ at
midrapidity
corresponding to  $p+p$ collisions at $\sqrt{s}=$13 TeV. From top to bottom, results are shown for
high-multiplicity,
low-multiplicity, and minimum bias events, respectively.}
\label{fig:6}
\end{figure}

\begin{figure}[htp]
\includegraphics[width=0.49\textwidth]{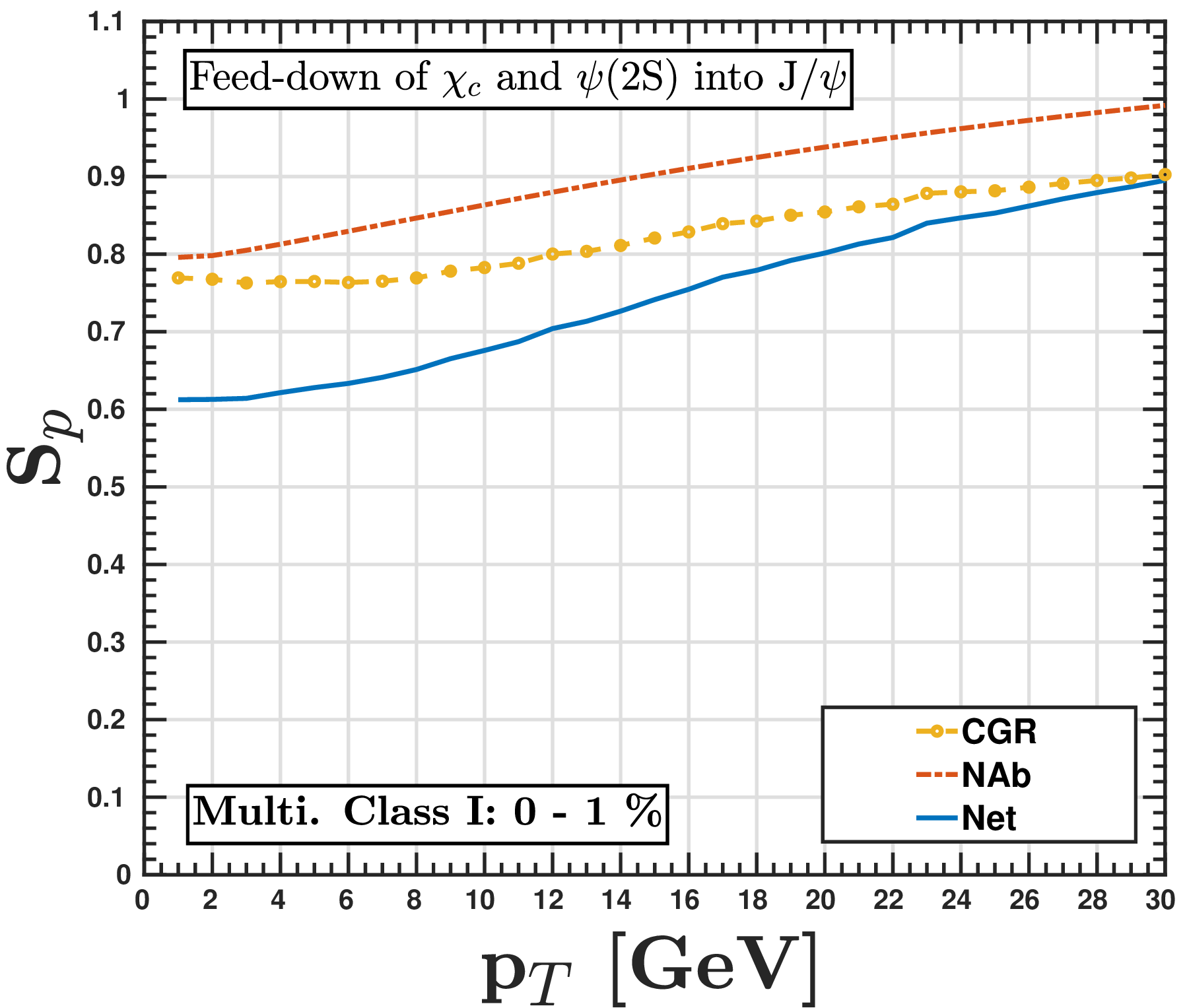}
\includegraphics[width=0.49\textwidth]{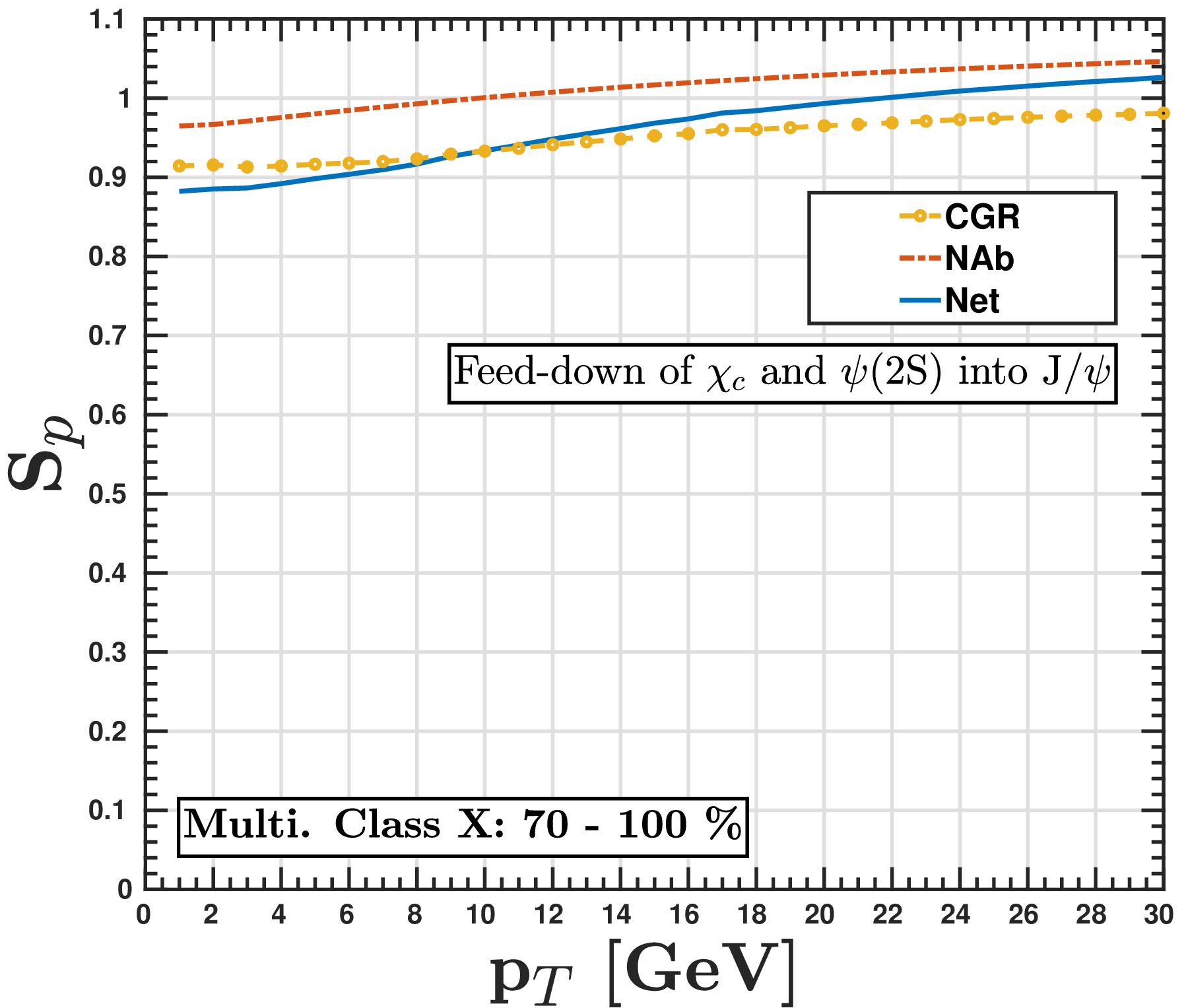}
\includegraphics[width=0.49\textwidth]{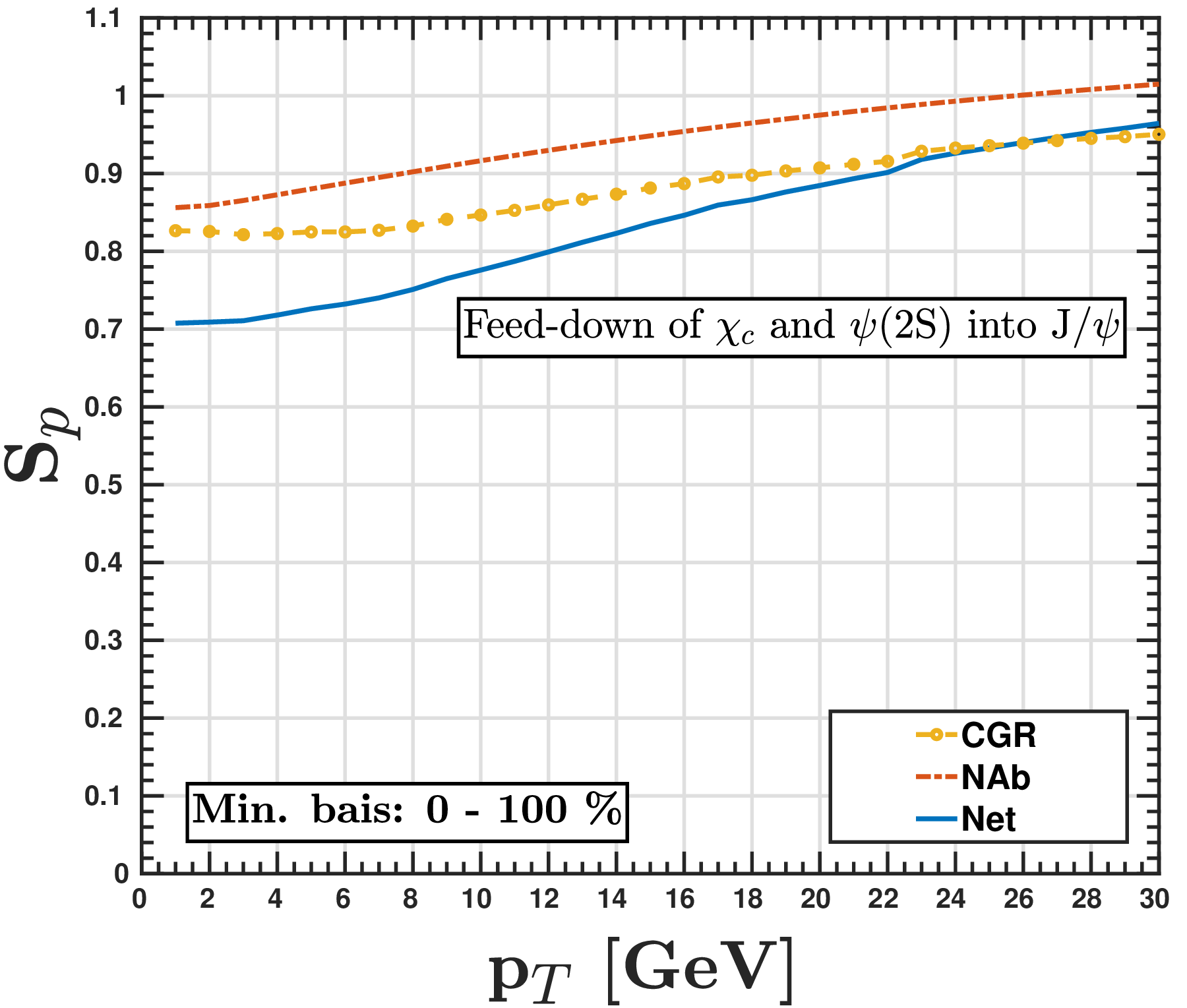}
\caption{(Color online) Survival probability $S_{P}$ as a function of $p_{T}$ is shown for $J/\psi$ considering
the feed-down of $\chi_{c}$(1P) and $\psi$(2S) into $J/\psi$ at midrapidity corresponding to  $p+p$ collision
at
$\sqrt{s}=$13 TeV. From top to bottom, results are shown for high-multiplicity, low-multiplicity, and minimum
bias events,
respectively.}
\label{fig:feed}
\end{figure}

\begin{figure}[htp]
\includegraphics[width=0.49\textwidth]{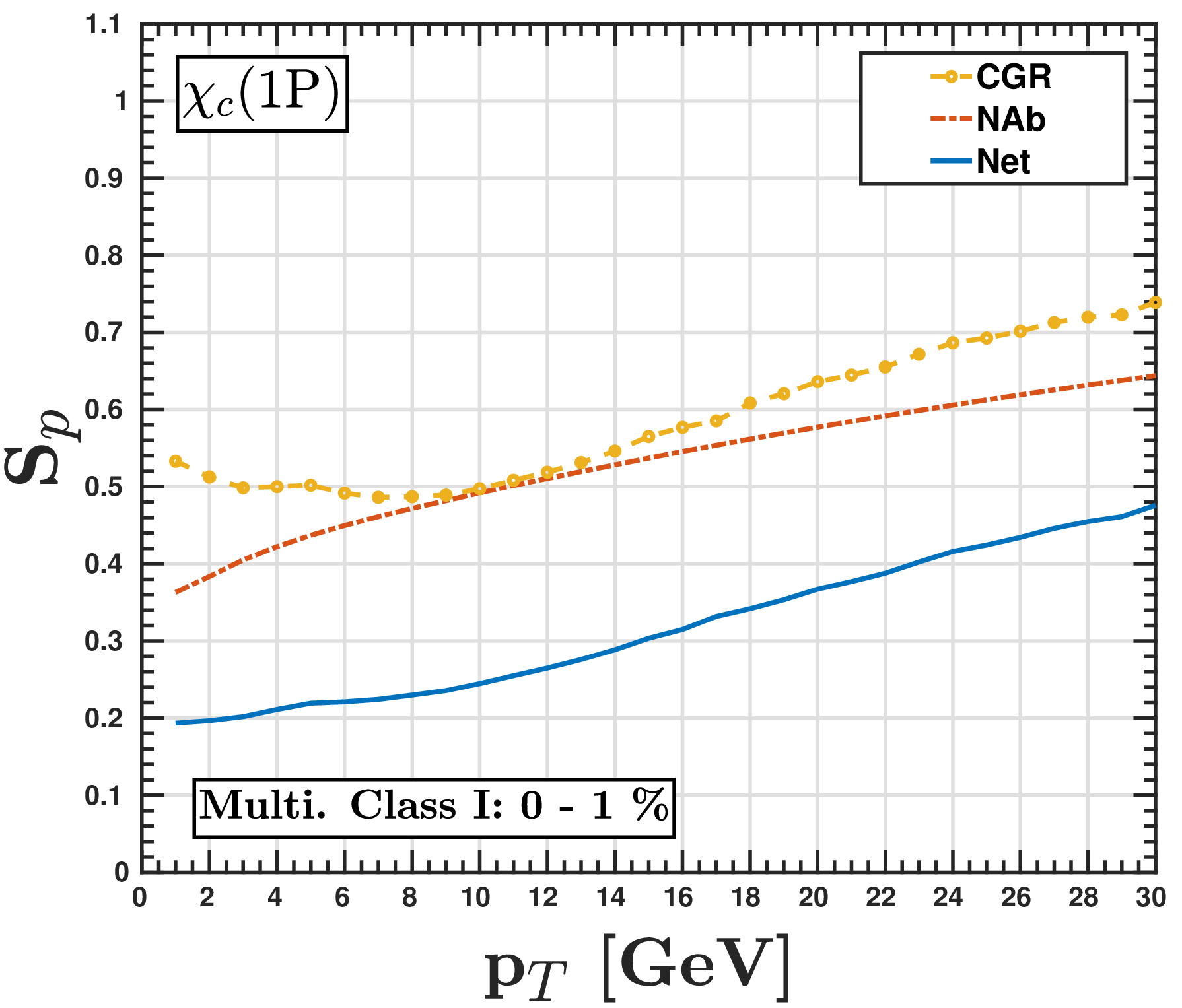}
\includegraphics[width=0.49\textwidth]{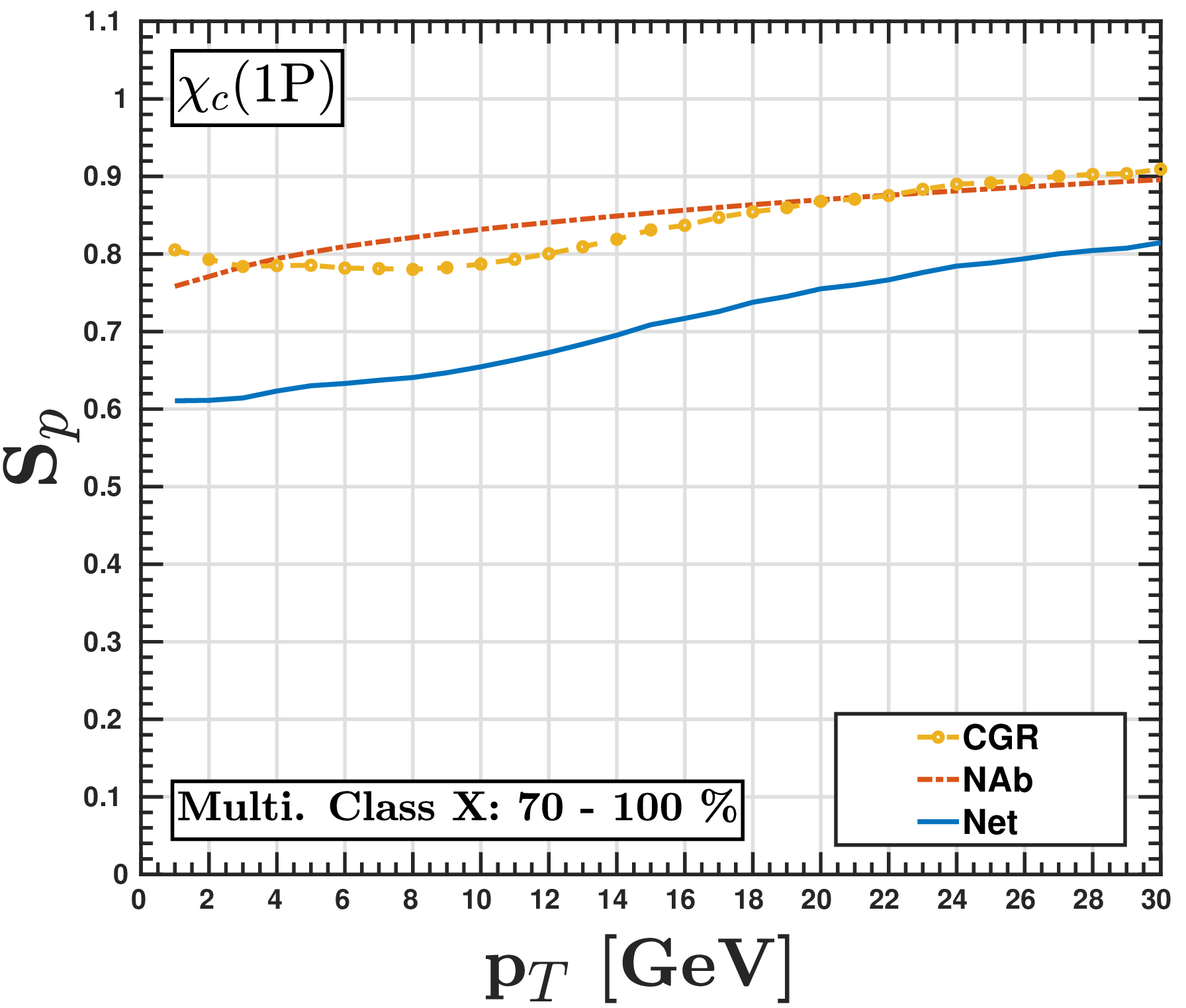}
\includegraphics[width=0.49\textwidth]{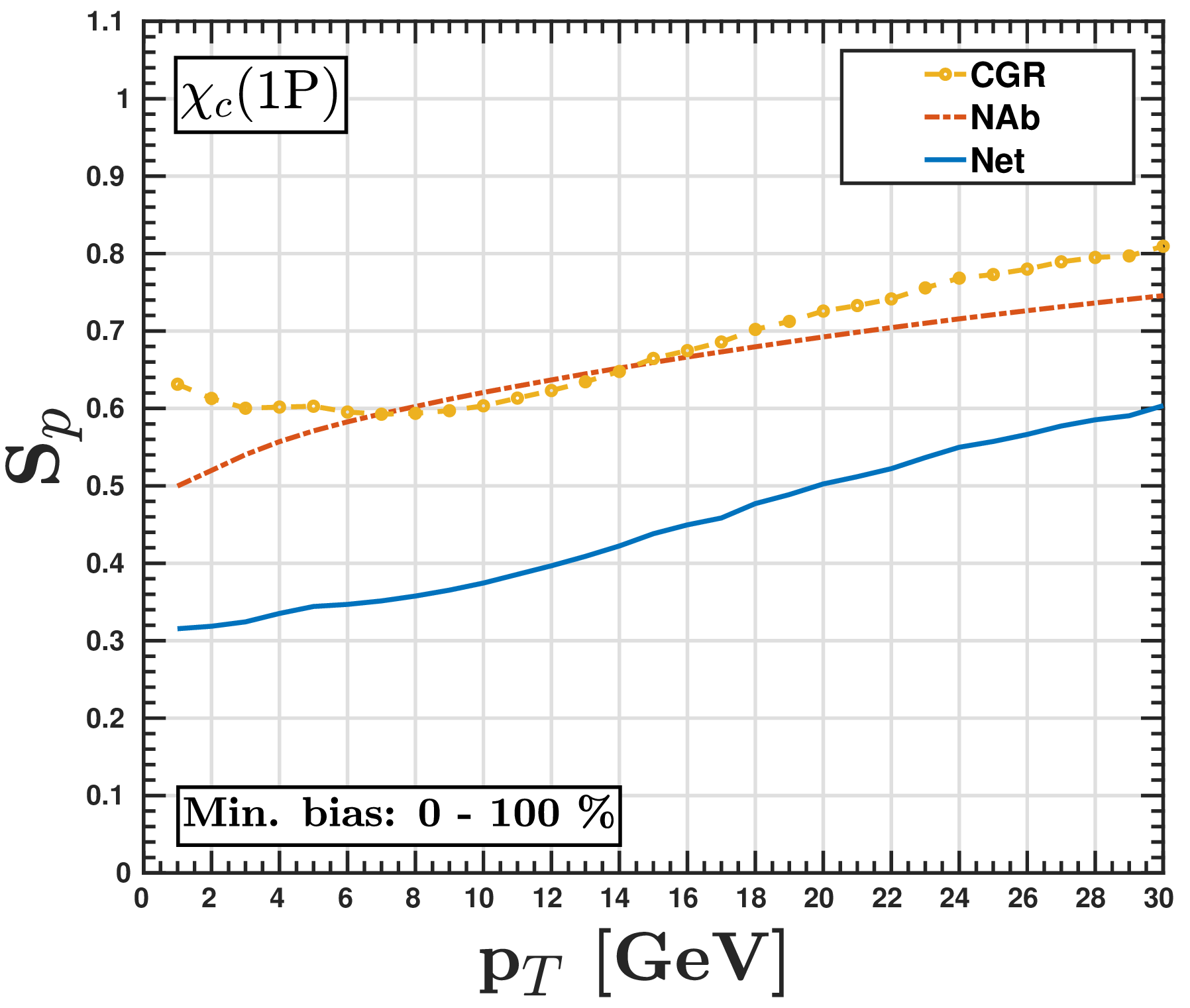}
\caption{(Color online) Survival probability $S_{P}$ as a function of $p_{T}$ is shown for $\chi_c$ at
midrapidity
corresponding to  $p+p$ collision at $\sqrt{s}=$13 TeV. From top to bottom, results are shown for
high-multiplicity, low-multiplicity,
and minimum bias events, respectively.}
\label{fig:7}
\end{figure}

The $p_{T}$-dependent suppression of $J/\psi$ corresponds to Multi. Class I, shown in Fig.~\ref{fig:6},
predicts
around 20\% suppression at low $p_{T}$ for both NAb and CGR mechanisms. However, as $p_{T}$ increases, the
suppression
due to NAb rapidly decreases and becomes negligible at $p_{T} \gtrsim 30$ GeV. On the other hand, the
suppression due
to
CGR also decreases with increasing $p_{T}$, but at a slower rate, still predicting around 10\% suppression at
$p_{T}
\simeq 30$ GeV. When these two mechanisms are combined, the suppression increases to around 40\% at low
$p_{T}$ and
10\% at high $p_{T}$, primarily due to the CGR mechanism. Fig.~\ref{fig:feed} shows that considering the
feed-down
corrections of $\chi_{c}$(1P) and $\psi$(2S) into $J/\psi$ at Multi. Class I provides marginal changes in the
results
compared to the
case without the feed-down. It suggests that high-multiplicity $\chi_{c}$(1P) is largely suppressed, and as
$\psi$(2S)
contribution in
feed-down is relatively small, mainly $J/\psi$ dynamics in the medium dominate the feed-down correction. The
results
for Multi.
Class X, depicted in Fig.~\ref{fig:6}, indicates a significant decrease in suppression due to NAb, with its
impact on
$J/\psi$ suppression being smaller than that of CGR at $p_{T} \lesssim 3$ GeV. The yield modification caused
by CGR is
also reduced at the lowest multiplicity and around $p_{T} \simeq 30$ GeV, where its effect nearly vanishes,
while NAb
deactivates at $p_{T} \simeq 30$ GeV. When these mechanisms are combined, they predict less than 10\%
suppression at low $p_{T}$,
which almost disappears at high $p_{T} \gtrsim 30$  GeV. However, the feed-down correction shown in
Fig.~\ref{fig:feed} for Multi.
Class X predicts a nonzero suppression at high $p_{T}$ due to CGR stemming from the larger suppression of
higher resonances.
In contrast, the corresponding NAb effect provides  a slight enhancement for $J/\psi$ at high $p_{T}$. In the
minimum bias case
illustrated in Fig. \ref{fig:6}, the prediction indicates a  significant suppression of $J/\psi$. When both
mechanisms are combined,
this suppression is approximately 30\% at $p_{T} \lesssim 3$ GeV, decreasing to 5\% at higher $p_{T}$. The
feed-down correction
shown in Fig.~\ref{fig:feed} for the minimum bias case slightly enhances the suppression for $J/\psi$. In this
case, the enhancement
of $J/\psi$ is found to be absent at high $p_{T}$, unlike Multi. Class X.\\

\begin{figure}[htp]
\includegraphics[width=0.49\textwidth]{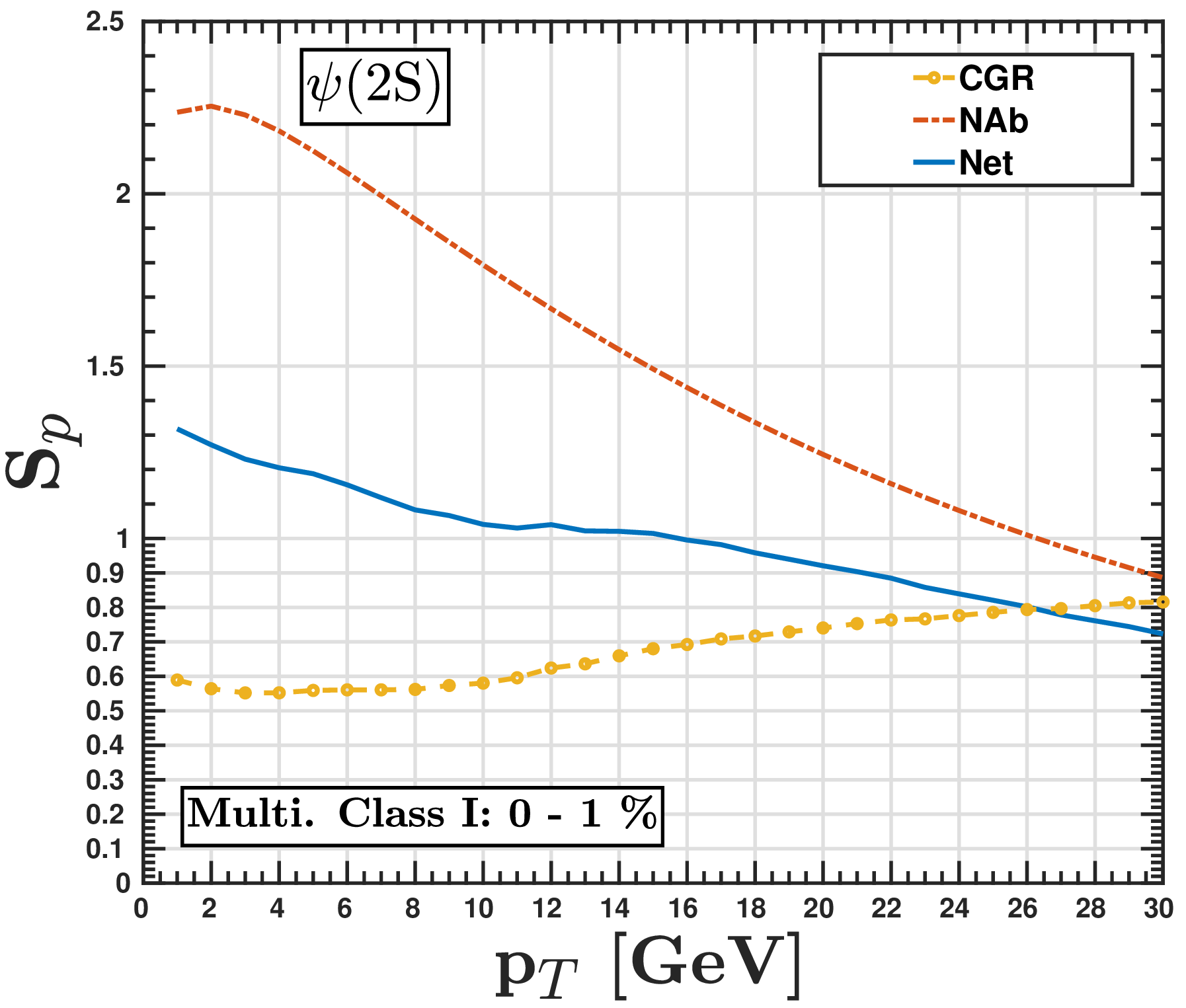}
\includegraphics[width=0.49\textwidth]{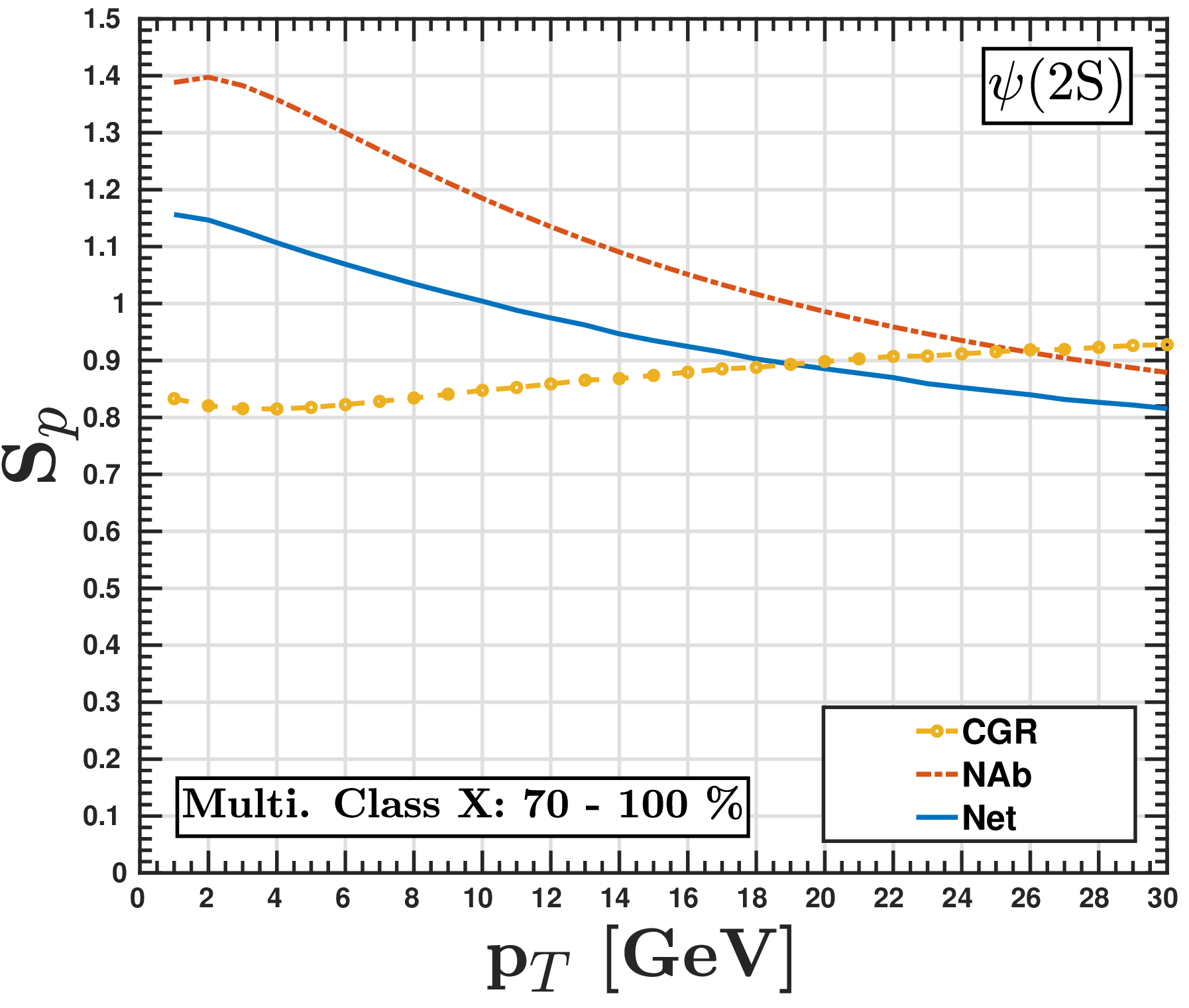}
\includegraphics[width=0.49\textwidth]{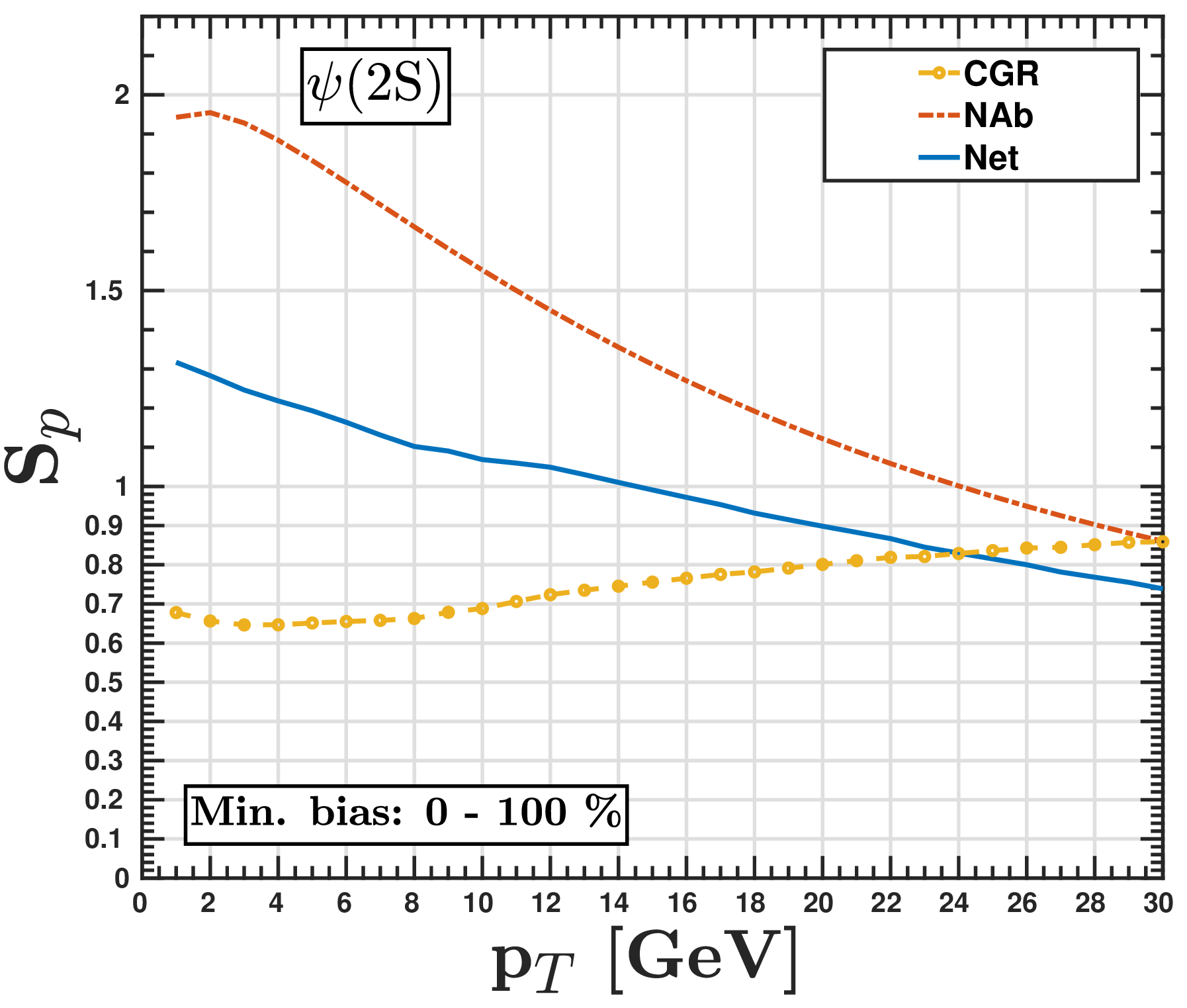}
\caption{(Color online) Survival probability $S_{P}$ as a function of $p_{T}$ is shown for $\psi$(2S) at
midrapidity
corresponding to  $p+p$ collision at $\sqrt{s}=$13 TeV. From top to bottom, results are shown for
high-multiplicity, low-multiplicity,
and minimum bias events, respectively.}
\label{fig:8}
\end{figure}

These results indicate that the nonadiabatic (NAb) evolution of the $J/\psi$ state is predominant at low
transverse
momentum ($p_{T}$) and high-multiplicity. At these conditions, the system size is maximized compared to lower
multiplicities, and particles with low $p_{T}$ moving slowly through the medium, prolong the transition from
$\psi(\tau_c)$ to $J/\psi$ at $\tau = \tau_{c}$. As a result, the yield of $J/\psi$ decreases under high
multiplicity
and low $p_{T}$, however, this reduction diminishes rapidly with increasing $p_{T}$ and a decrease in system
size.\\

In parallel, the combined effects of CGR
significantly
influence suppression. Within CGR, collisional damping and gluonic dissociation substantially lower the yield
at low
$p_{T}$, whereas the regeneration mechanism tends to increase the yield at high $p_{T}$, rendering suppression
less
impactful for $J/\psi$. Conversely, for other charmonium resonances like $\chi_{c}$(1P) and $\psi$(2S), the
regeneration effect is marginal. Given their excited state nature, these resonances experience greater
suppression
compared to $J/\psi$ due to CGR mechanisms.\\

Interestingly, NAb affects these states differently. For example, in Multi. Class I, $\chi_{c}$(1P) exhibits
dominant
suppression due to the NAb mechanism as shown in Fig~\ref{fig:7}. This suppression results in a yield
reduction of
approximately 65\% at $p_{T} \lesssim 2$, while survival probability at high $p_{T}$ increases to about 55\%.
In
contrast, the CGR suppression for $\chi_{c}$(1P) in Fig~\ref{fig:7} at $p_{T} \lesssim 2$ is around 45\%,
rising to
50\% in the range of $2 \lesssim p_{T} \lesssim 12$ GeV before declining at $p_{T} \gtrsim 12$ GeV. The
interplay of CGR
and NAb predicts a suppression range of 50\% to 80\% for $\chi_{c}$(1P) across high to low $p_{T}$ in high
multiplicity
(Multi. Class I) events.\\

Further, Fig.~\ref{fig:7} illustrates for low-multiplicity events (Multi. Class X), the CGR and NAb mechanisms
exhibit a complex relationship regarding $\chi_{c}$(1P) suppression from low to high $p_{T}$. At $p_{T}
\lesssim 3$
GeV, NAb is the primary suppression mechanism; however, in the range of $3 < p_{T} < 18$ GeV, dissociation of
$\chi_{c}$(1P) is largely driven by CGR processes. At high transverse momenta ($p_{T} \gtrsim 20$), CGR and NAb
equally affect the yield of $\chi_{c}$(1P) in low-multiplicity events, resulting in net suppression between
20\% and
40\%, depending on the $p_{T}$ region. Similar to Multi. Class I, NAb primarily drives suppression mechanisms
for
$\chi_{c}$(1P) in a minimum bias scenario, except in the $p_{T}$ range of 8 - 14 GeV, where CGR predicts
greater
suppression. The overall suppression of $\chi_{c}$(1P) in minimum bias (Min. bias: 0 - 100\%) lies between the
extremes
of multiplicity classes, ranging from approximately 70\% suppression at low $p_{T}$ to about 40\% at high
$p_{T}$.\\

So far, observations indicate that nonadiabatic evolution tends to reduce the yields of quarkonia, as has been
predicted for $J/\psi$ and $\chi_c$. However, the results shown in Fig.~\ref{fig:8} for $\psi$(2S) is on the
contrary. Instead of suppression, nonadiabatic evolution leads to a significant enhancement of $\psi$(2S)
yields
across both low and high-multiplicity classes. In Multi. Class I, Fig.~\ref{fig:8} reveals a substantial
enhancement of
$\psi$(2S) at low transverse momentum ($p_{T}$), which diminishes as $p_{T}$ increases. At very high $p_{T}$
(around 26
GeV and above), there is a noticeable suppression pattern for $\psi$(2S). A similar phenomenon is observed in
Multi.
Class X, though the magnitude of enhancement is smaller compared to Class I, with the yield of $\psi$(2S)
starting to
decrease at $p_{T} \gtrsim 10$ GeV.\\

These findings suggest that the nonadiabatic evolution of charmonium states facilitates the transition to
excited
states characterized by larger principal quantum numbers ($n$) and smaller azimuthal quantum numbers ($l$).
This
transition is particularly dominant when the lifetime of the medium is sufficiently long, allowing the
continuum state
to evolve into a discrete charmonium state. Given that $\psi$(2S) is a higher excited state with relatively
high eigenenergy, it is particularly conducive to the formation during this transition from continuum to
discrete eigenstates.
Consequently, the nonadiabatic mechanism predicts a significant enhancement of $\psi$(2S) yields at high
multiplicity,
which then decreases due to changes in eigenenergy in lower-multiplicity events.\\

On the other hand, the CGR mechanisms significantly reduce the yield of $\psi$(2S) across all chosen
multiplicity
classes, as illustrated in Fig.~\ref{fig:8}. For Multi. Class I, the suppression is around 4$-$45\% at $p_{T}
\lesssim
$12 GeV, further with increasing $p_{T}$ suppression reduces to 20\%. The combined effects of CGR and NAb lead
to a
decrease in the ``Net" survival probability ($\rm S_{p}$) for $\psi$(2S) at high $p_{T}$, which in contrast
with the
behavior observed for $J/\psi$ and $\chi_c$. A similar trend reflects the influence of NAb and CGR on the
$\psi$(2S) yield with  $p_{T}$, is also seen in Multiplicity Class X; however, the magnitude of enhancement and
suppression is less pronounced due to change in charged-particle multiplicity density. Under the  Min. bias
scenario,
the $\psi$(2S) yield falls within the range set by Multi. Class I and Multi. Class X and results are
consistent with
the behavior observed in other multiplicity classes.\\

\begin{figure*}[htp]
\includegraphics[width=0.49\textwidth]{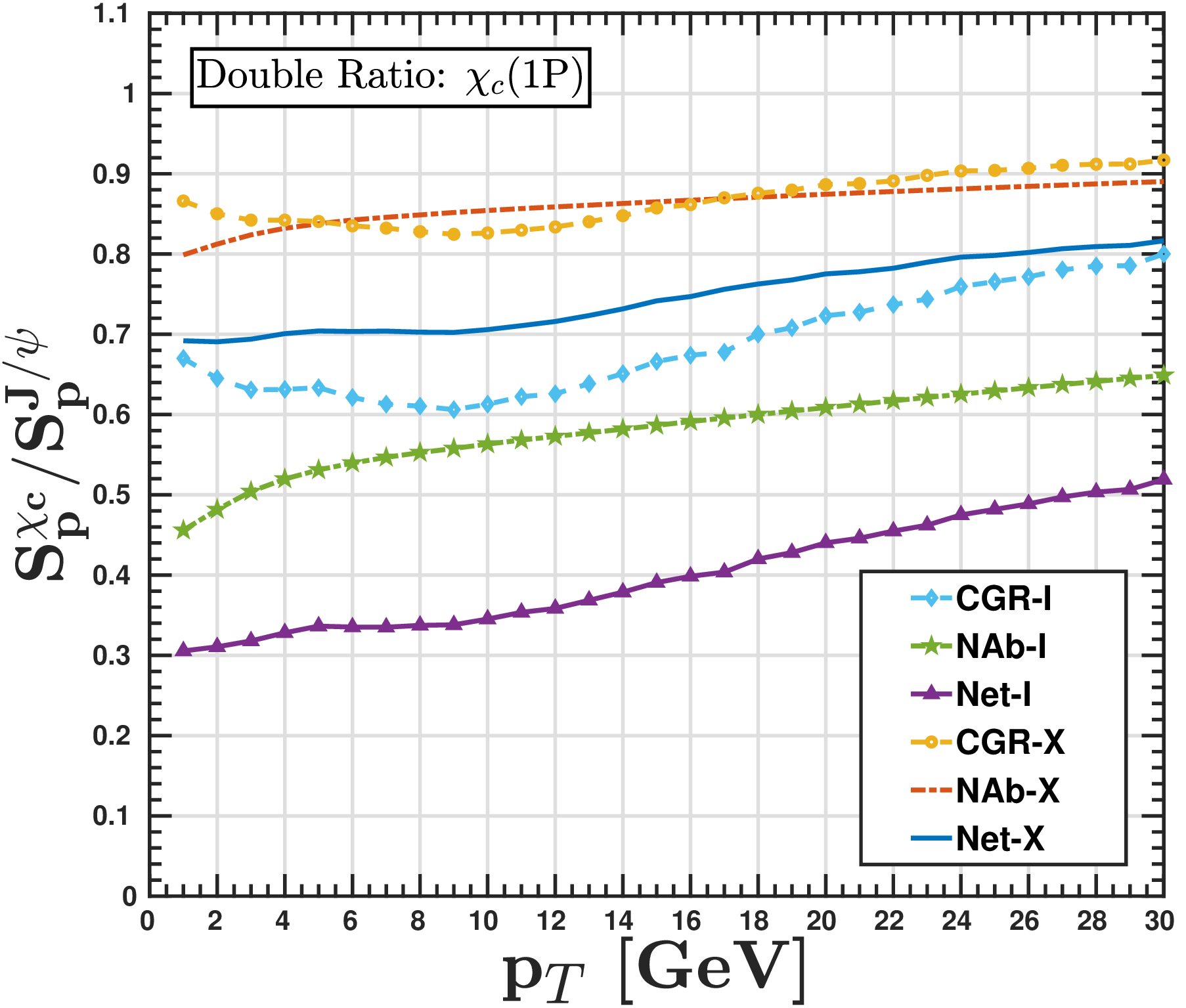}
\includegraphics[width=0.49\textwidth]{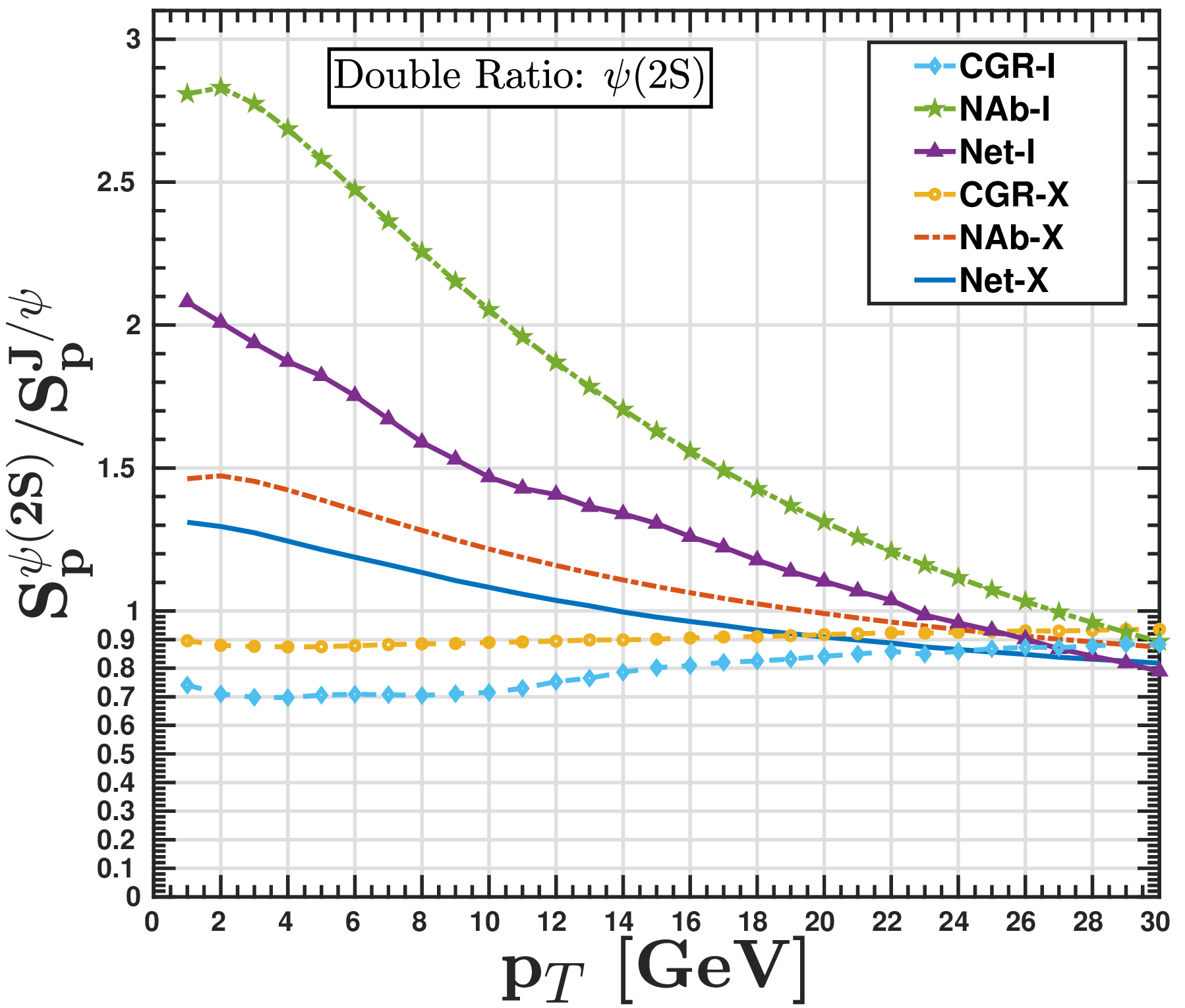}
\caption{(Color online) Double ratio as a function of $p_{T}$ is shown for $\frac{\chi_{c}(1P)}{J/\psi}$,  and
$\frac{\psi(2S)}{J/\psi}$,  at midrapidity corresponding to  $p+p$ collision at $\sqrt{s}=$13 TeV. Legends
shown with
``I'' and ``X'', stand for High and Low multiplicity events, respectively.}
\label{fig:9}
\end{figure*}

\begin{figure*}[htp]
\includegraphics[width=0.49\textwidth]{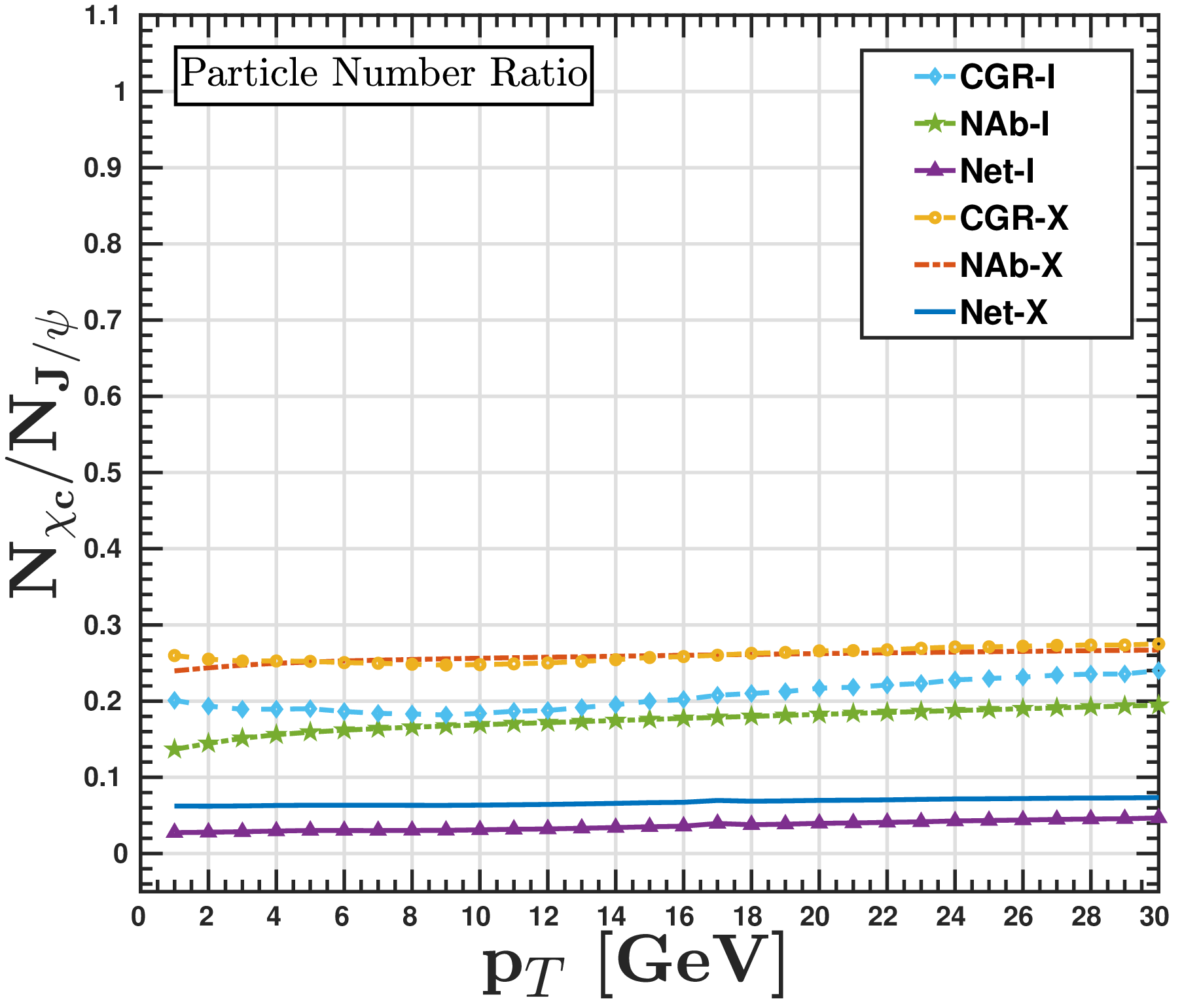}
\includegraphics[width=0.49\textwidth]{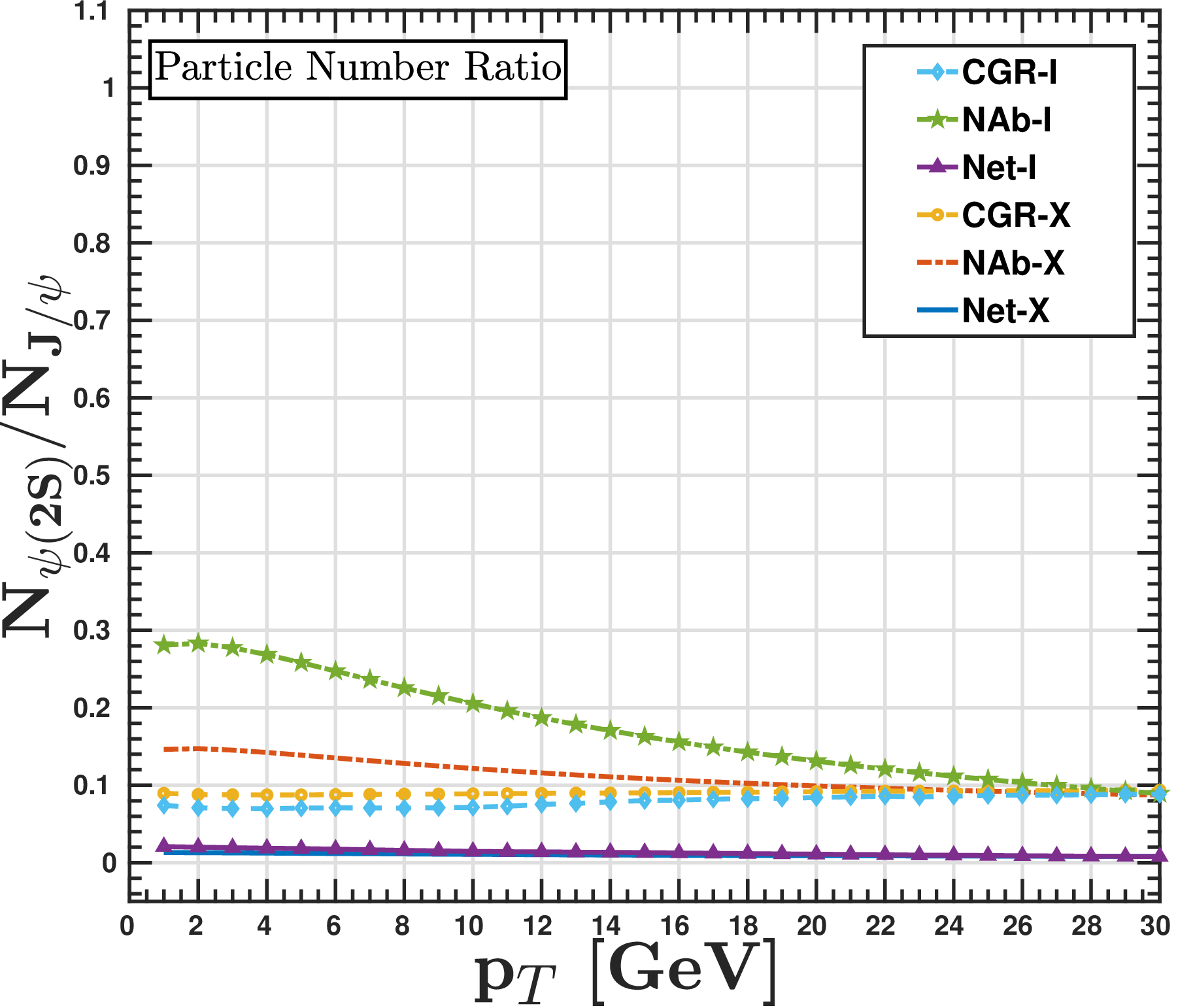}
\caption{(Color online) Particle number ratio as a function of $p_{T}$ is shown for
$\frac{\chi_{c}(1P)}{J/\psi}$,  and
$\frac{\psi(2S)}{J/\psi}$,  at midrapidity corresponding to  $p+p$ collision at $\sqrt{s}=$13 TeV. Legends
shown with
``I'' and ``X'', stand for High and Low multiplicity events, respectively.}
\label{fig:10}
\end{figure*}

The $p_{T}$-dependent double ratios depicted in Fig.~\ref{fig:9} provide important insights into the relative
suppression of $\chi_{c}$(1P) and $\psi$(2S) in comparison to $J/\psi$. At low-multiplicity and $p_{T}$, the
NAb
mechanism demonstrates a higher degree of suppression than the CGR mechanism for $\chi_{c}$(1P) relative to
$J/\psi$. Notably, at
$p_{T} \gtrsim 4$ GeV, the suppression for $\chi_{c}$(1P) predicted by both mechanisms align closely,
indicating a
convergence in their outcomes under certain conditions.\\

In high-multiplicity scenarios, the NAb mechanism is identified as a significant factor in the reduction of the
$\chi_{c}$(1P) yield, leading to a suppression approximately 60\% to 50\% greater than that of $J/\psi$ across
the
selected $p_{T}$ range. While the CGR mechanism predicts substantial suppression for $\chi_{c}$(1P), its
estimates are notably
lower than the NAb mechanism. Collectively, these findings suggest a net suppression of $\chi_{c}$(1P) with
respect to
$J/\psi$  is around 70\% to 50\% at high-multiplicity and from 30\% to 20\% at low-multiplicity within the
$p_{T}$ range of $1
\leq p_{T} \leq 30$ GeV.\\

Moreover, the left panel of Fig.~\ref{fig:9} illustrates that the relative yield of $\psi$(2S) experiences a
suppression of approximately 10\% at low-multiplicity and 30\% at high-multiplicity due to the CGR mechanism.
Particularly, the
$\psi$(2S) considerable enhancement through the NAb mechanism underscores the distinct roles played by each
mechanism
in particle dynamics. The survival probability of $\psi$(2S) in comparison with $J/\psi$ increases at both low
and high
multiplicities. However, it is noteworthy that at $p_{T} \gtrsim 14$ GeV in low-multiplicity scenarios,
$\psi$(2S) is
observed to be more suppressed than $J/\psi$. In higher-multiplicity conditions, the onset of suppression for
$\psi$(2S) is shifted to higher $p_{T}$ values, as evidenced by results indicating suppression at $p_{T}
\gtrsim 22$ GeV. This
subtle understanding enhances our comprehension of particle behavior across varying system dynamics depending
on the
charged-particle multiplicity density.\\

We have conclusively observed that the $\chi_c$ survival probability ($\rm{S_P}$) in the QGP medium is
significantly
lower than that of $J/\psi$ and $\psi$(2S). While $\psi$(2S) is indeed suppressed due to the CGR mechanisms,
that too
is  less pronounced than that of $\chi_c$. Additionally, $\psi$(2S) experiences enhancement from the  NAb
mechanism. This
leads us to critical questions about whether the net production of $\psi$(2S) actually exceeds that of
$J/\psi$ or if
its survival probability is merely bolstered under these conditions.\\

To resolve this, we have estimated the relative production numbers of $\chi_c$ and $\psi$(2S) with respect to
$J/\psi$,
as illustrated in Fig.~\ref{fig:10} through particle number ratios. At low-multiplicity, the ratio of $\chi_c$
to
$J/\psi$ shows that approximately 30\% of the final production of $\chi_c$ is conceded when factoring in CGR
and NAb
mechanisms independently. Notably, this final yield is nearly independent of the transverse momentum ($p_T$)
of the
particles. The combined effects of CGR and NAb definitively reduce the net $\chi_c$ yield by up to 8\% at low
multiplicity. At high-multiplicity, the production of $\chi_c$ under CGR mechanisms accounts for around 20\%
to 25\%,
whereas predictions based solely on NAb yield estimates of 12\% to 20\%, fluctuating from low to high $p_T$.
The
cumulative impacts of CGR and NAb conspicuously diminish the net $\chi_c$ production by approximately 2\% in
high
multiplicity events.\\

Furthermore, when examining the net yield of $\psi$(2S) relative to $J/\psi$, it is found that the final yield
of
$\psi$(2S) is substantially lower than that of $J/\psi$. At high $p_T$, the production levels of $\psi$(2S) are
comparable to those of $\chi_c$, driven by the NAb mechanism. However, at low $p_T$, the yield of $\psi$(2S) is
significantly less than that of $\chi_c$ across both multiplicity classes. With CGR mechanisms taken into
account, the
yield of $\psi$(2S) is estimated at approximately 8\% to 10\% relative to $J/\psi$ for both high and low
multiplicity.
The combined effects of CGR and NAb indicate that the production of $\psi$(2S) is roughly 1\% of $J/\psi$ and
which is
smaller than $\chi_c$. This trend remains consistent across low and high-multiplicity as well as throughout the
selected $p_T$ ranges.\\

These results strongly suggest that while the sequential suppression of charmonium may appear
inconsistent in this context, the sequential production of charmonium states is upheld. Even when accounting
for the
complexities of medium dynamics and charmonium evolution in ultrarelativistic $p+p$ collisions, it is evident
that
$\psi$(2S) may experience enhancements; however, the net number of $J/\psi$ will invariably surpass that of
$\psi$(2S).

\section{Summary and Outlook}
\label{summary}

This work explored the charmonium yield modification under various mechanisms that could possibly exist in
ultrarelativistic proton-proton ($p+p$) collisions at $\sqrt{s}$ = 13 TeV. The study considers both
preequilibrium
and thermalized QCD medium effects, modeling the temperature evolution through the bottom-up thermalization
approach
and Gubser flow. Under the ``in-medium suppression effects" for QGP, it incorporates collisional damping,
which arises
because of the energy loss due to interactions between charmonium and the medium, and gluonic dissociation as
the
consequence of quarkonium states into a color octet lead interactions with gluons. It also includes the
regeneration of
charmonium states within the medium due to the transition from the color octet state to the color singlet
state.
Additionally, the nonadiabatic evolution of charmonium states is considered, recognizing that rapid
temperature
changes in small systems like $p+p$ collisions can challenge the adiabatic assumption and significantly affect
the
charmonium yield. At last, feed-down corrections from higher resonances into $J/\psi$ have been incorporated
for more
realistic predictions.\\

\begin{itemize}
\item The findings conclude that charmonium suppression is driven by these mechanisms, and their combined
effect is
modeled in terms of survival probabilities ($\rm S_{P}$) as a function of transverse momentum ($p_T$) and
charged-particle multiplicity ($dN_{ch}/d\eta$). The study finds that while  $J/\psi$ and $\chi_c$ experience
significant suppression, $\psi$(2S) shows enhancement at higher multiplicities due to nonadiabatic evolution
at low
$p_{T}$ and high multiplicities.

\item These results indicate that the QGP evolution timescale is significantly smaller than the charmonium
transition
timescale in ultrarelativistic $p+p$  collisions, thereby invalidating the use of the adiabatic approximation
for the
state evolution in the medium. This discrepancy necessitates considering a nonadiabatic evolution of
charmonium,
especially in small systems such as those formed in ultrarelativistic $p+p$ and even in ultraperipheral
heavy-ion
collisions.

\item The results suggest that $J/\psi$ suppression and/or $\psi$(2S) enhancement in small systems, such as
$p+p$
collisions can be a valuable probe for understanding the presence of a thermalized QCD medium. This
investigation
suggests that ultrarelativistic $p+p$ collisions may also exhibit QGP-like behavior under specific conditions.
\end{itemize}
This study presented a holistic approach that reinforces our understanding of quark-gluon plasma
characteristics and
enhances our grasp of the intricate dynamics within ultrarelativistic collisions from large to small systems.\\

The future scope of research on charmonium yield modification in $p+p$ collisions at $\sqrt{s} = 13$ TeV
offers various
promising directions:

\begin{itemize}
 \item {\it Non-Adiabatic Evolution:} The study shows that the evolution of quarkonia in smaller systems, such
as $p+p$
collisions, may not adhere to adiabatic assumptions. Future work can further explore the nonadiabatic
evolution of
charmonium states, particularly in different system sizes, like peripheral heavy-ion collisions, where rapid
cooling
influences their behavior.

\item {\it Comparison with Heavy-Ion Collisions:} Our findings are contrary to the $\psi$(2S) suppression
observed in heavy-ion collisions and that controversy arises because the evolution of charmonium states is
considered
to behave differently depending on the system size and its cooling rate. Future research should investigate
this
phenomenon more deeply to understand how charmonium states behave across various collision systems.

\item {\it QGP characteristics in small systems:} The findings suggest the potential for using charmonium
suppression as
a probe to detect thermalized QCD matter, even in small systems like $p+p$ collisions. Further experimental
studies
could focus on developing the methodology for such observations to probe the existence of quark-gluon plasma
in such a
small collision system.

\end{itemize}

These avenues can help connect theory with experimental observations, enhancing the understanding of QGP
properties and
charmonium dynamics in ultrarelativistic nuclear and/or hadrons collisions.

\section*{Acknowledgment}
This research work has been conducted with financial support from the DAE-DST, Government of India, as part of
the Mega-Science
Project ``Indian Participation in the ALICE experiment at CERN" with
Project
No. SR/MF/PS-02/2021-IITI (E-37123). Captain R. Singh and Raghunath Sahoo acknowledge the financial support
received
from the aforementioned DAE-DST Project. Partha Bagchi acknowledges financial support under the DAE project
No. RIN 4001.



\end{document}